\documentclass[%
reprint,
superscriptaddress,
showpacs,preprintnumbers,
 amsmath,amssymb,
 aps,
prb,
floatfix,
]{revtex4-1}

\usepackage{epsfig}
\usepackage[thinspace]{SIunits}
\usepackage{xspace}
\usepackage{graphicx}
\usepackage{dcolumn}
\usepackage{bm}
\usepackage{color}
\usepackage{hyperref}
\hypersetup{
    unicode=false,          
    pdftoolbar=true,        
    pdfmenubar=true,        
    pdffitwindow=false,     
    pdfstartview={FitH},    
    pdftitle={My title},    
    pdfauthor={Author},     
    pdfsubject={Subject},   
    pdfcreator={Creator},   
    pdfproducer={Producer}, 
    pdfkeywords={keyword1} {key2} {key3}, 
    pdfnewwindow=true,      
    colorlinks=true,       
    linkcolor=blue,          
    citecolor=blue,        
    filecolor=blue,      
    urlcolor=blue,       
    plainpages=false        
}

\begin{document}

\title{Superconductivity and fluctuation in Ba$_{1-p}$K$_p$Fe$_2$As$_2$ and Ba(Fe$_{1-n}$Co$_n$)$_2$As$_2$}
\date{\today}
\author{T.~B\"ohm}
\affiliation{Walther Meissner Institut, Bayerische Akademie der Wissenschaften, 85748 Garching, Germany}
\affiliation{Fakult\"at f\"ur Physik E23, Technische Universit\"at M\"unchen, 85748 Garching, Germany}
\author{R.~Hosseinian~Ahangharnejhad}
\affiliation{Walther Meissner Institut, Bayerische Akademie der Wissenschaften, 85748 Garching, Germany}
\affiliation{Fakult\"at f\"ur Physik E23, Technische Universit\"at M\"unchen, 85748 Garching, Germany}
\author{D. Jost}
\affiliation{Walther Meissner Institut, Bayerische Akademie der Wissenschaften, 85748 Garching, Germany}
\affiliation{Fakult\"at f\"ur Physik E23, Technische Universit\"at M\"unchen, 85748 Garching, Germany}
\author{A. Baum}
\affiliation{Walther Meissner Institut, Bayerische Akademie der Wissenschaften, 85748 Garching, Germany}
\affiliation{Fakult\"at f\"ur Physik E23, Technische Universit\"at M\"unchen, 85748 Garching, Germany}
\author{B. Muschler}
\altaffiliation{Present address: Zoller \& Fr\"ohlich GmbH, Simoniusstrasse 22, 88239 Wangen im Allg\"au, Germany}
\affiliation{Walther Meissner Institut, Bayerische Akademie der Wissenschaften, 85748 Garching, Germany}
\affiliation{Fakult\"at f\"ur Physik E23, Technische Universit\"at M\"unchen, 85748 Garching, Germany}
\author{F.~Kretzschmar}
\altaffiliation{Present address: Intel Mobile Communications, Am Campeon 10-12, 85579 Neubiberg, Germany}
\affiliation{Walther Meissner Institut, Bayerische Akademie der Wissenschaften, 85748 Garching, Germany}
\affiliation{Fakult\"at f\"ur Physik E23, Technische Universit\"at M\"unchen, 85748 Garching, Germany}
\author{P. Adelmann}
\affiliation{Karlsruher Institut f\"ur Technologie, Institut f\"ur Festk\"orperphysik, 76021 Karlsruhe, Germany}
\author{T. Wolf}
\affiliation{Karlsruher Institut f\"ur Technologie, Institut f\"ur Festk\"orperphysik, 76021 Karlsruhe, Germany}
\author{Hai-Hu Wen}
\affiliation{National Laboratory of Solid State Microstructures and Department of Physics, Nanjing University, Nanjing 210093, China}
\author{J.-H. Chu}
\affiliation{Stanford Institute for Materials and Energy Sciences,
SLAC National Accelerator Laboratory, 2575 Sand Hill Road, Menlo Park, CA 94025, USA}
\affiliation{Geballe Laboratory for Advanced Materials \& Dept. of Applied Physics,
Stanford University, CA 94305, USA}
\author{I.~R. Fisher}
\affiliation{Stanford Institute for Materials and Energy Sciences,
SLAC National Accelerator Laboratory, 2575 Sand Hill Road, Menlo Park, CA 94025, USA}
\affiliation{Geballe Laboratory for Advanced Materials \& Dept. of Applied Physics,
Stanford University, CA 94305, USA}
\author{R. Hackl}
\affiliation{Walther Meissner Institut, Bayerische Akademie der Wissenschaften, 85748 Garching, Germany}

\begin{abstract}
  We study the interplay of fluctuations and superconductivity in BaFe$_2$As$_2$ (Ba-122) compounds with Ba and Fe substituted by K ($p$ doping) and Co ($n$ doping), respectively. To this end we measured electronic Raman spectra as a function of polarisation and temperature. We observe gap excitations and fluctuations for all doping levels studied. The response from fluctuations is much stronger for Co substitution and, according to the selection rules and the temperature dependence, originates from the exchange of two critical spin fluctuations with characteristic wave vectors $(\pm\pi, 0)$ and $(0,\pm\pi)$. At 22\% K doping ($p=0.22$), we find the same selection rules and spectral shape for the fluctuations but the intensity is smaller by a factor of 5. Since there exists no nematic region above the orthorhombic spin-density-wave (SDW) phase the identification of the fluctuations via the temperature dependence is not possible. The gap excitations in the superconducting state indicate strongly anisotropic near-nodal gaps for Co substitution which make the observation of collective modes difficult. {The variation with doping of the spectral weights of the $A_{1g}$ and $B_{1g}$ gap features does not support the influence of fluctuations on Cooper pairing. Therefore, the observation of Bardasis-Schrieffer modes inside the nearly clean gaps on the K-doped side remains the only experimental evidence for the relevance of  fluctuations for pairing.}
\end{abstract}
  
\pacs{74.70.Xa, 
      74.20.Mn, 
      74.25.nd, 
      74.40.-n  
     }
\maketitle

\section{Introduction}
Since the seminal study of multi-band superconductors by Suhl, Matthias and Walker \cite{Suhl:1959} it is known that the combination of either intra- and inter-band or of different types of interactions can lead to substantially enhanced superconducting transition temperatures $T_c$ \cite{Carbotte:1990,Johnston:2010a}. The relevance of these considerations was demonstrated
for monolayer FeSe on SrTiO$_3$. \cite{GeJF:2015,LeeJJ:2014} In addition to experimental studies there are recent theoretical predictions as to the enhancement of $T_c$ through fluctuations of the charge, orbital or spin degrees of freedom \cite{Lederer:2015}. The question is how one can demonstrate the influence of various pairing mechanisms and how ideas as to the realization of materials with higher $T_c$ values can be developed.

It is obvious that a single spectroscopic method such as Raman scattering cannot pin down one or more routes to Cooper pairing and disentangle their individual influence. However, it has been demonstrated that one can get an  idea which interactions can contribute \cite{Muschler:2010,Caprara:2011} or establish a hierarchy of interactions \cite{Kretzschmar:2013,Bohm:2014}. In addition, light scattering affords a window into the fluctuations above various phase transitions \cite{Caprara:2005,Tassini:2005,Eiter:2013} which may contribute to the pairing in the superconducting state \cite{Lederer:2015,Caprara:2011,Perali:1996,Gallais:2015,Kretzschmar:2016}. One essential advantage of light scattering is the existence of selection rules. Beyond the well known selection rules for phonons or spin excitations one may discriminate between electronic excitations in different regions of the Brillouin zone (BZ) \cite{Devereaux:1994,Mazin:2010a} or project out excitations or critical fluctuations with characteristic wave vectors ${\bf q}_c$ \cite{Muschler:2010,Caprara:2011,Caprara:2005}.

Exploiting these selection rules several open issues in the iron-based compounds could be addressed \cite{Kretzschmar:2013,Bohm:2014,Gallais:2015,Kretzschmar:2016,Muschler:2009,Chauviere:2010,Thorsmolle:2016}. It was observed that the superconducting energy gap $2\Delta_i(\bf{k})$ depends on band $i$ and momentum {\bf k} and even may have near nodes in Ba(Fe$_{1-n}$Co$_n$)$_2$As$_2$ (BFCA) close to optimal doping ($n\approx 0.06$) \cite{Muschler:2009}. In hole-doped Ba$_{1-p}$K$_p$Fe$_2$As$_2$ (BKFA) the gap is still band dependent but shows little variation on the individual bands \cite{Kretzschmar:2013} in agreement with angle-resolved photoemission spectroscopy (ARPES) \cite{Evtushinsky:2009}. In addition, there are strong indications of two nearly degenerate pairing channels \cite{Bohm:2014} suggesting interband pairing and unconventional  coupling \cite{Scalapino:2009,Scalapino:2012}. The question arises as to the underlying fluctuations. In fact, fluctuations were clearly observed in parent BaFe$_2$As$_2$ and electron doped Ba(Fe$_{1-n}$Co$_n$)$_2$As$_2$ for $n \lesssim 0.08$ \cite{Kretzschmar:2016,Choi:2010,Gallais:2013} but not yet in Ba$_{1-p}$K$_p$Fe$_2$As$_2$.

In this paper we present data on Ba$_{1-p}$K$_p$Fe$_2$As$_2$ for $p=0.22$ and demonstrate that the fluctuations can be identified also here. We compare the pure fluctuation response with the results from the thermodynamic measurements. In addition, we show spectra in the superconducting state of Ba(Fe$_{1-n}$Co$_n$)$_2$As$_2$ at all main in-plane symmetries for $0.041\le n \le 0.085$ and find that the intensities below $T_c$ exhibit maxima at optimal doping in both $A_{1g}$ and $B_{1g}$ symmetry.

\begin{figure}[tbp]
  \centering
  \includegraphics[width=1.0\columnwidth]{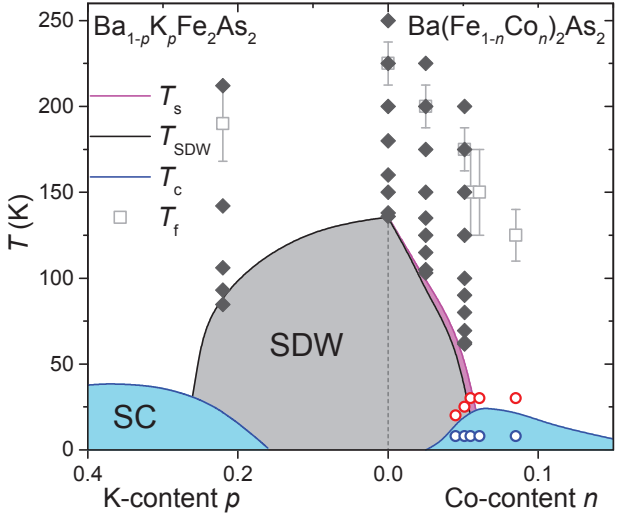}
  \caption{Phase diagram of hole and electron-doped BaFe$_2$As$_2$, as adopted from Ref. \cite{Bohmer:2014}. The grey diamonds indicate the positions of the Raman experiments above and at the magneto-structural transition with the data on the electron-doped side reproduced from Ref.~\cite{Kretzschmar:2016}. The onset temperature of fluctuations $T_f$ is given by open squares. The open circles represent the data of BFCA above and below $T_c$. The magenta wedge indicates the nematic phase.
  }
  \label{fig:cartoon}
\end{figure}

\section{Experiment}
The single crystals of hole-doped Ba$_{1-p}$K$_p$Fe$_2$As$_2$ and electron-doped Ba(Fe$_{1-n}$Co$_n$)$_2$As$_2$ were grown using a self-flux technique and have been characterized elsewhere \cite{Shen:2011,Karkin:2014,Chu:2009}. The concentrations of K and Co were determined by microprobe analysis. For the Raman measurements samples with narrow superconducting transitions were selected having $\Delta T_c$ values in the range 0.4 to 2\,K. The doping levels and typical sample temperatures are displayed in Fig.~\ref{fig:cartoon}.

The experiments were performed with standard light scattering equipment. For excitation a diode-pumped solid state lasers (Coherent Genesis MX\,SLM; Klastech Scherzo-DENICAFC-532-300) and  an Ar$^+$ (Coherent Innova {304}) laser were used emitting at 575, 532 and 514.5\,nm, respectively.  The samples were mounted on the cold finger of a He-flow cryostat in a cryogenically pumped vacuum. The laser-induced heating was determined experimentally to be close to 1\,K per mW absorbed power. The majority of the spectra was measured only in the three polarization configurations $xy$, $x^\prime y^\prime$, and $RR$ where $x$ and $y$ refer to Fe-Fe bonds and $x^\prime=1/\sqrt{2}(x+y)$, $y^\prime=1/\sqrt{2}(y-x)$, $R  =1/\sqrt{2}(x+iy)$. For the symmetry assignment in the 1\,Fe unit cell, which we use throughout this paper, these polarizations project the (electronic) $B_{2g}+A_{2g}$, $B_{1g}+A_{2g}$, and $A_{1g}+A_{2g}$ symmetries, respectively. We found that the $A_{2g}$ contributions can be ignored since they are temperature independent and typically smaller than 20\% of those of the other symmetries in the energy range studied here.

The spectra shown below represent the imaginary part of the Raman susceptibiity $R\chi^{\prime\prime}(\Omega,T)$ which is obtained by dividing the  cross section by the Bose thermal factor $\{1+n(T,\Omega)\}=[1-\exp(-\hbar\Omega/k_BT)]^{-1}$; $R$ is an experimental constant. In the $B_{1g}$  spectra we isolate the contribution from critical fluctuations by subtracting the electron-hole (e-h) continuum which is found to follow the real part {of the} optical conductivity as $\Omega\chi^{\prime\prime}(\Omega,T) \propto \sigma^\prime(\Omega,T)$ in agreement with theoretical predictions \cite{Shastry:1990}.

\section{Results}
\subsection{BKFA}
\begin{figure}[tbp]
  \centering
  \includegraphics[width=1.0\columnwidth]{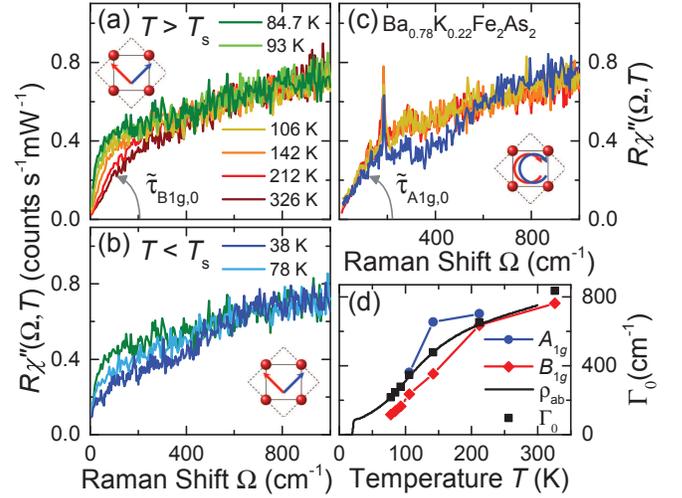}
  \caption{Raman spectra of Ba$_{1-p}$K$_p$Fe$_2$As$_2$ for $p=0.22$ above the superconducting transition $T_c$. The polarizations are indicated schematically. (a), (b) Data in $B_{1g}$ symmetry above and below the magneto-structural transition at $T_s=81.4$\,K. (c) $A_{1g}$ spectra. The opening of a gap due to SDW order can be observed in the 38\,K spectra of both $A_{1g}$ and $ B_{1g}$ symmetry. (d) Transport and static Raman relaxation rates. The resistivity \cite{Shen:2011} is shown as a black line after conversion into a relaxation rate.
  }
  \label{fig:BKFA}
\end{figure}
$A_{1g}$ and $B_{1g}$ spectra of ${\rm Ba_{0.78}K_{0.22}Fe_2As_2}$ are plotted in Fig.~\ref{fig:BKFA} (a)-(c). For $\Omega > 600\,{\rm cm}^{-1}$, the spectra and their variation with temperature are independent of symmetry. Similarly, we observe a suppression of the scattering intensity in the energy range below $600\,{\rm cm}^{-1}$ and a weak increase around $800\,{\rm cm}^{-1}$ in both symmetries below 85\,K [Fig.~\ref{fig:BKFA}\,(b) and (c)] which originates in the formation of the gap in the SDW phase. For $\Omega < 400\,{\rm cm}^{-1}$  the temperature dependence in $B_{1g}$ symmetry is much stronger than in $A_{1g}$ symmetry. This difference becomes particularly clear in the analysis of the initial slope, $\tilde{\tau}_{\mu,0}(T) = R\lim_{\Omega\to 0}\left[\chi^{\prime\prime}_\mu(\Omega,T)/\Omega\right]$ ($\mu = B_{1g}, ~A_{1g}$) as defined in Fig.~\ref{fig:BKFA}\,(a) and (c). $\tilde{\tau}_{\mu,0}(T)$ includes the unknown intensity factor $R$ that relates the slope and the Raman relaxation time $\tau_{\mu,0}(T)$. In Fig.~\ref{fig:BKFA}\,(d) we show the corresponding static relaxation rates $\Gamma_{\mu,0}(T)=\hbar/\tau_{\mu,0}(T)$, that can  be derived in absolute energy units \cite{Opel:2000}, and compare it with the results derived from the resistivity \cite{Shen:2011}. The $B_{1g}$ results follow the temperature dependence of the resistivity above 220\,K but vary much stronger in the range 85--220\,K.

We interpret this enhanced variation in terms of a new scattering channel opening up below approximately 220\,K due to fluctuations and analyze the data similarly as in the case of underdoped Ba(Fe$_{1-n}$Co$_n$)$_2$As$_2$ \cite{Kretzschmar:2016}. For extracting the response of the fluctuations we subtract the electron-hole (e-h) continuum from the total response. The e-h continuum is approximated in a way that the spectra above 220\,K are fully reproduced. Below 220\,K we vary the e-h continuum slightly with temperature by adjusting the parameters appropriately to reproduce the intensity above $800\,{\rm cm}^{-1}$ and make the initial slope to follow the resistivity. The resulting relaxation rates are shown as black squares in Fig.~\ref{fig:BKFA}\,(d). More details can be found in the Supplementary Information of Ref.~\cite{Kretzschmar:2016}.
\begin{figure}[tbp]
  \centering
  \includegraphics[width=1.0\columnwidth]{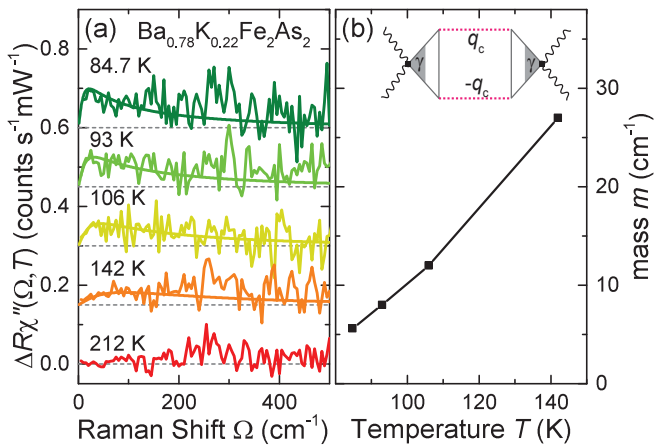}
  \caption{Fluctuation contribution to the Raman spectra of Ba$_{1-p}$K$_p$Fe$_2$As$_2$ for $p=0.22$. (a) Experimental spectra and theoretical prediction \cite{Caprara:2005}. The spectra are shifted for clarity. The respective zero of intensity is indicated by a dashed line. (b) Temperature dependence of the mass $m(T)$ of the propagator. {The inset shows the type of diagrams used for the analysis \cite{Caprara:2005,Kretzschmar:2016}. Wavy, solid and dashed lines represent photons, electrons, and critical fluctuations, respectively.}
  }
  \label{fig:BKFA_fluct}
\end{figure}

The results of the fluctuation response are presented in Fig.~\ref{fig:BKFA_fluct}. In addition to the experimental data we show theoretical predictions on the basis of Aslamazov-Larkin diagrams that describe the exchange of two critical fluctuations with finite but opposite momenta $\pm{\bf q}_c$ \cite{Caprara:2005}. The theory does not \textit{a priori} specify the origin of the propagators and can equally well be used for spin, charge or orbital fluctuations. Small corrections apply if the propagator couples to the lattice \cite{Caprara:2005}. As in the case of the cuprates \cite{Tassini:2005} or of BFCA \cite{Kretzschmar:2016} quantitative agreement between experiment and theory is found for realistic parameters. In particular, the intensity and the mass of the fluctuation propagator $m(T)\propto \xi^{-2}$, with $\xi$ the correlation length, are determined at one temperature, and the response at the other temperatures is reproduced by just varying $m(T)$. In comparison to BFCA, the overall intensity is smaller by a factor of five either as a result of a resonance effect in BFCA or of weaker fluctuations in a material without a nematic phase. However, as will be shown below, $m(T)$ and the variation with temperature are comparable.
\begin{figure}[tbp]
  \centering
  \includegraphics[width=1.0\columnwidth]{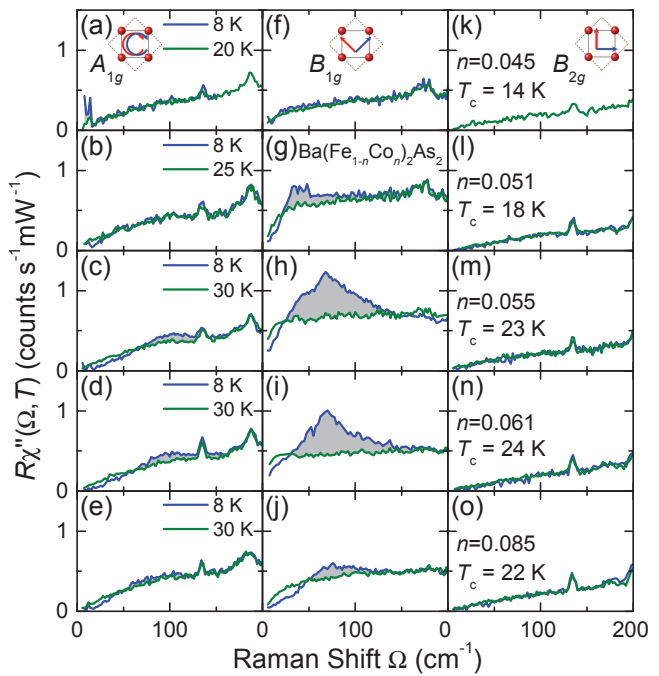}
  \caption{Raman spectra of BFCA as a function of doping $n$. $n$ and $T_c$ are given in the right column. The data were taken above and below $T_c$ as indicated (see also Fig.~\ref{fig:cartoon}) at all three main main polarization configurations. Since the $A_{2g}$ spectra have negligibly small intensity \cite{Muschler:2009}, the data in the three columns show essentially the $A_{1g}$, $B_{1g}$, and $B_{2g}$ symmetries. The spectral weights in the pair breaking peaks are indicated in grey.
  }
  \label{fig:BFCA}
\end{figure}

\subsection{BFCA}
In Fig.~\ref{fig:BFCA} we show the doping dependence of the superconducting spectra of BFCA in all symmetry projections including $A_{1g}$ which was not studied before \cite{Chauviere:2010,Gallais:2013}. The figure shows spectra right above and well below $T_c$ at temperatures indicated in the first column of Fig.~\ref{fig:BFCA} and by open circles in Fig.~\ref{fig:cartoon}. The e-h continua observed above $T_c$ (green) are similar in all symmetries. However, there is a strong doping dependence in $B_{1g}$ symmetry (second column) in that the initial slope becomes very steep around optimal doping ($n=0.55$ and 0.61). In the superconducting state (blue) there is a redistribution of intensity from low to high energies in $A_{1g}$ and $B_{1g}$ symmetry leading to a reduction below typically 20-40\,cm$^{-1}$ due to the energy gap and an enhancement in the range 50-150\,cm$^{-1}$ originating in pair breaking and excitations across the gap (grey area). At optimal doping, we observe a square-root-like increase of the intensity in the low-energy part of the superconducting $B_{1g}$ spectra [Fig.~\ref{fig:BFCA}\,(i)] which was interpreted in terms of accidental nodes on the Fermi surface of the outer electron band and a high density of states (DOS) inside the gap \cite{Mazin:2010a,Muschler:2009}.

In addition to the high DOS there are secondary structures at around 50\,cm$^{-1}$ for $n=0.55$ and 0.61 [Fig.~\ref{fig:BFCA}\,(h) and (i)]. We do not believe that they originate from band-dependent gaps \cite{Chauviere:2010}. Rather, they may be remainders of a collective Bardasis-Schrieffer mode similar to that observed in BKFA \cite{Bohm:2014} which are barely visible because of quasi-particle damping in a material with a strongly momentum dependent gap on a single band \cite{Devereaux:1995a} and the concomitant high DOS below the gap maximum.

\section{Discussion}
We address now the fluctuations in the normal state above the magnetically ordered phase and superconductivity and discuss the interrelation of fluctuations, possible nematic order and superconducting pairing.

\subsection{Nematicity and electronic Raman scattering}
The Raman response, in contrast to the optical conductivity, does not obey the usual $f$-sum rule \cite{Kosztin:1991,Freericks:2005}, and several scattering channels can open up as a function of, for instance, temperature. In some cases the response from different channels is just additive such as for weakly coupled phonons. They are superposed on the e-h continuum which, then, reflects symmetry-resolved transport properties \cite{Shastry:1990,Opel:2000,Zawadowski:1990,Devereaux:2003,Devereaux:2007}. For strongly coupled phonons the response from charge and lattice has to be treated on equal footing, and the line shape assumes an asymmetric Fano-type energy dependence \cite{Opel:1999b}. In a strongly coupled superconductor normal and superconducting response approach each other at an energy of several times the maximal gap $\Delta_0$ and are interrelated in a complicated way at low energies \cite{Manske:2004} as can be seen in Fig.~\ref{fig:BFCA}.

Contributions to the response from critical fluctuations \cite{Caprara:2005,Tassini:2005,Kretzschmar:2016,Choi:2010,Gallais:2013,%
Thorsmolle:2016,Yoon:2000,Venturini:2002c,Gallais:2016} can be either superposed on the e-h continuum \cite{Caprara:2005,Khodas:2015,Karahasanovic:2015,Yamase:2013,Yamase:2015} or develop out of it \cite{Gallais:2016}. In both cases the related susceptibility and the integrated spectral weight become critical upon approaching the phase transition and diverge in the limit $\Omega=0$. If the fluctuations interact among each other and/or couple to the lattice a phase transition can be induced before the susceptibility diverges \cite{Fernandes:2012}.

In the Fe-based systems there are various types of instabilities which can drive phase transitions. Since all systems have magnetic phases one may conclude that spin fluctuations are the leading instability. However, depending on the sign of the interaction between the hole bands in the center of the Brillouin zone and the electron bands around $(\pm\pi,0)$ and $(0,\pm\pi)$ also orbital/charge fluctuations can dominate \cite{Kontani:2010,Fernandes:2014}. For addressing this problem, Kretzschmar and coworkers studied BFCA where the magnetic ordering temperature $T_{\rm SDW}$ and the structural transition $T_s>T_{\rm SDW}$ are separated \cite{Kretzschmar:2016}. In the nematic phase between $T_{\rm SDW}$ and $T_s$ which has orbital but no magnetic order fluctuations can still be observed arguing for spin rather than charge fluctuations which are expected to disappear at $T_s$. If the spin fluctuations interact among themselves, where $g_0$ describes the electron-mediated interaction, the light couples to the electronic nematic susceptibility $\chi^{\rm el}_{\rm nem,0}(T)$ which is driven by the spin susceptibility $\chi_{\rm mag}(q)$ as \cite{Fernandes:2014}
\begin{equation}
  \label{eq:chinemel}
  \chi^{\rm el}_{\rm nem,0}(T)=\frac{\int_{q}  \chi_{\rm mag}^2(q)}{1-g_0\int_{q} \chi_{\rm mag}^2(q)}.
\end{equation}
The magnetic susceptibility diverges at $T_{\rm SDW}$. For $g_0\ge0$ $\chi^{\rm el}_{\rm nem,0}(T)$ has a Curie-like $|T-T_0|^{-1}$ divergence at $T_0\ge T_{\rm SDW}$. Close to $\Omega=0$ the Raman response of interacting spin fluctuations, $R\tilde{\chi}^{\prime\prime}_{\rm f}(\Omega,T)$, is given by \cite{Kretzschmar:2016}
\begin{equation}
  \label{eq:raman}
  R\tilde{\chi}^{\prime\prime}_{\rm f}(\Omega,T)= R{\chi}^{\prime\prime}_{\rm f}(\Omega,T) \left [ 1+g_0 \chi^{\rm el}_{\rm nem,0} (T) \right ].
\end{equation}
${\chi}^{\prime\prime}_{\rm f}(\Omega,T)$ describes the line shape of non-interacting fluctuations \cite{Caprara:2005}, $\chi^{\rm el}_{\rm nem,0} (T)$ accounts for the variation of the intensity. Eq.~\eqref{eq:raman} is valid only for small energies, and the initial slope of the spectra is proportional to the variation of the spectral weight.

Finally, the presence of magneto-elastic coupling shifts the structural phase transition to higher temperature, $T_s>T_0$, since the coupling $g_0$ will be renormalized as ${g}=g_0+\lambda^2/C^2_{66,0}$ where $C_{66,0}\approx 40$\,GPa \cite{Yoshizawa:2012} is the nearly doping-in\-de\-pen\-dent high-temperature limiting value of the shear modulus $C_{66}(T)$ and $\lambda$ is the coupling constant in the bilinear term of the Landau free energy density. As a consequence, the spectral weight does not diverge at $T_s>T_0$ but has only a maximum.

The analysis of the BFCA data supports the spin nematic scenario \cite{Kretzschmar:2016} and may even indicate an interrelation between fluctuations and superconductivity \cite{Gallais:2015}. The latter proposal is a particular motivation for studying fluctuations in BKFA, having the highest $T_c$ in the BFA family, and for a more detailed look at the evolution with doping of the superconducting spectra of BFCA.

\subsection{Fluctuations and doping}
BKFA does not have a nematic phase as BFCA. In addition, the magnetic and structural transitions coincide rendering the phase transformation first order. Upon comparing the Raman results on the fluctuations one finds the intensity in BKFA (see Fig.~\ref{fig:BKFA_fluct}) to be much smaller than in BFCA \cite{Kretzschmar:2016}. Arguably, the intensity is not a good quantity in a light scattering experiment. However, the overall intensity in the $B_{1g}$ channel has little doping and material dependence, as can be read directly from Figs.~\ref{fig:BKFA} and \ref{fig:BFCA}, and does not show strong resonances \cite{Mazin:2010a}. In contrast, one observes a huge intensity variation of the fluctuation response close to $T_s$.

\begin{figure}[tbp]
  \centering
  \includegraphics[width=0.99\columnwidth]{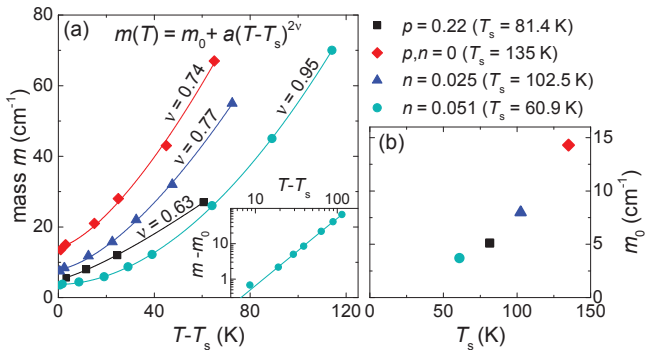}
  \caption{{Temperature and doping dependence of the mass $m(x,T)$ ($x=p,n$) in the fluctuation propagator. The masses are derived from the fluctuation contribution to the Raman spectra according to the analysis of Ref.~\cite{Caprara:2005}. The upper right part of the figure shows the correspondence between samples and symbols. The structural transition temperatures are indicated. (a) The variation of the mass follows a power law with critical exponent $\nu$. The inset on the lower right displays $m(x=0,T)$ on a log-log scale. $\nu$ depends on doping (Fig.~\ref{fig:parameters1}). (b) The offset of the mass $m_0$ varies monotonically with $T_s$ and by and large extrapolates to zero for $T_s \to 0$.}
  }
  \label{fig:mass}
\end{figure}

In contrast to the intensities, the relaxation rates (see Fig.~\ref{fig:BKFA} and Ref.~\cite{Opel:2000}) and the masses of the fluctuation propagator (see Fig.~\ref{fig:BKFA_fluct} and Ref.~\cite{Caprara:2005}) can be derived in absolute units. Therefore we start by comparing the masses
\begin{equation}
  \label{eq:mass}
  m(x,T) = m_0(x)+a(x)|T-T_s(x)|^{2\nu(x)}
\end{equation}
for the four doping levels available at the moment, where $x=n,p$. Fig.~\ref{fig:mass}\,(a) shows $m(x,T)$ for BKFA and BFCA. Fits to the data using Eq.~\eqref{eq:mass} yield the offset $m_0(x)$ close to $T_s$ and the critical exponent $\nu(x)$ which depends substantially on doping $x$. The inset shows on a log-log scale that the data are indeed well described by a power low. Fig.~\ref{fig:mass}\,(b) displays the the variation of $m_0$. We find the mass to decrease monotonously with $T_s(x)$ without a significant influence of the type of substitution. Although data for samples with lower $T_s$ would be desirable one can observe the trend of $m_0(T_s)$ to vanish linearly with $T_s$. Hence, in the limit $T_s\to0$  $m(x,T)$ is expected to become scale free as predicted for a quantum critical point (QCP). Since $T_s$ vanishes on either side of zero doping the mass of the fluctuation propagator suggests the existence of two QCPs in agreement with other methods. We note that $m\to0$ is equivalent to a diverging correlation length $\xi$ or an ordered phase with vanishing transition temperature $T_s$ in accordance with the definition of a QCP.

\begin{figure}[tbp]
  \centering
  \includegraphics[width=0.90\columnwidth]{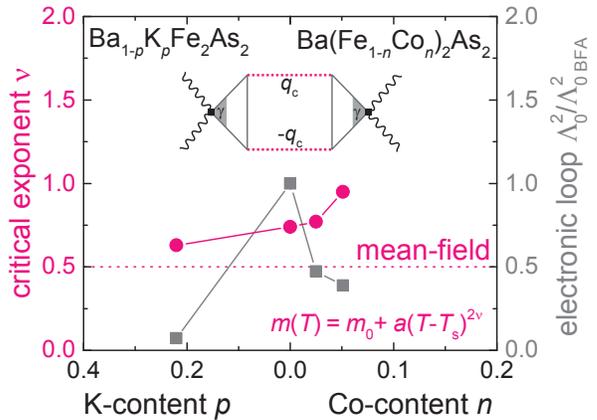}
  \caption{{Electronic loop $\Lambda_0^2$ of the AL diagrams and critical exponent $\nu$ of the mass $m(T)$ as a function of doping $x=p,n$. $\Lambda_0^2$ (triangles in the diagram) is determined experimentally via a fit at one temperature and kept constant above $T_s$. It includes the effect of the Raman matrix elements. $\Lambda_0^2$ is strongly peaked at $n=p=0$ (BFA). The mass has a power-law dependence on $T$ [Eq.~\eqref{eq:mass}]. The exponent $\nu$ depends monotonically on doping and approaches the mean field value of 0.5 on the $p$-doped side.}
  }
  \label{fig:parameters1}
\end{figure}

The critical exponent $\nu$ depends monotonically on doping, as shown in Fig.~\ref{fig:parameters1}, and has the tendency to approach the value of 0.5, predicted in the meanfield approximation, for $p$-doped materials. For Co substitution ($n$-doping), $\nu$ increases towards the QCP ($n\approx0.06$) and reaches a value close to unity for the highest doping level studied here. Whereas $m_0(T_s)$ scales with $T_s$, one finds the critical exponent to scale with $x$. Currently, we do not have an explanation but can pinpoint an obvious $n-p$ asymmetry in the type of fluctuations.

Fig.~\ref{fig:parameters1} shows also the intensity prefactor $\Lambda_0^2$ as a function of doping. We remphasize that $\Lambda_0^2$ is a temperature independent electronic property. In the approach here, it depends on states close to the Fermi surface thus yielding the selection rules \cite{Caprara:2005}. However, it depends non-monotonically on both doping (see figure) or $T_s$ and does not follow the overall intensity of the e-h continuum. There are two possible explanations: (i) Resonance effects play a role and indicate different orbital selectivity for e-h excitations and for fluctuations. (ii) The relative shapes of the electron- and hole-like Fermi surfaces are doping dependent. The better the overlap the stronger the fluctuations \cite{Khodas:2015}. The latter scenario would indeed explain the maximum at $x=0$ where a relatively well-defined nesting vector, equivalent with a match of the Fermi surface shapes, induces a $(\pi,0)$ spin density wave instability. With increasing doping the nesting becomes worse (on either side) and the intensity $\Lambda_0^2$ decreases. One could then argue that the nesting deteriorates more rapidly on the hole-doped side thus driving the system further away from the SDW instability and expanding the Fermi surface available for superconductivity.

\subsection{Fluctuations and elastic constants}

Obviously fluctuations precede the SDW phase in general and exist at least up to doping levels at which superconductivity commences. First, the Raman response will be compared to the evolution of the elastic constants in BKFA, similarly as performed for BFCA \cite{Gallais:2013,Gallais:2016,Bohmer:2016}.

To this end the static electronic nematic susceptibility $\chi^{\rm el}_{\rm nem,0}(T)$ needs to be derived which, using Landau theory, was shown to govern the temperature dependence of the shear modulus $C_{66}$ \cite{Cano:2010,Fernandes:2010},
\begin{equation}
  \frac{C_{66}}{C_{66,0}}= 1 - \frac{\lambda^2}{C_{66,0}} \chi^{\rm el}_{\rm nem,0}(T).
  \label{eq:C66}
\end{equation}

The phase transition temperature $T_s$ is determined by $C_{66}\to 0$ or $\chi^{\rm el}_{\rm nem,0}(T)= C_{66,0}/\lambda^2$ hence above the divergence point $T_0$. If the phase transition is driven by an electronic instability $C_{66}$ does not necessarily need to go completely to zero \cite{Yamase:2013,Yamase:2015,Bohmer:2016}.

If the lattice phase transition couples to the electronic nematicity the (1\,Fe) $B_{1g}$ Raman response can couple to $C_{66}$. There are various ways to search for a possible coupling. In a first study the entire $B_{1g}$ spectra were analyzed \cite{Gallais:2013} by deriving the real part of the static Raman susceptibility $R\chi^{\prime}_{\rm B1g,0}(T)$ via Kramers-Kr\"onig (K-K) transformation from the experimental response $R\chi^{\prime\prime}_{B1g}(\Omega,T)$ and identifying $R\chi^{\prime}_{\rm B1g,0}(T)$ with $\chi^{\rm el}_{\rm nem,0}(T)$. In the limit $\Omega=0$ the K-K transform is identical to the first moment of $R\chi_{B1g}^{\prime\prime}(\Omega)/\Omega$,
\begin{equation}
  \tilde{R}_{B1g}(T)=\frac{2 R}{\pi}\int_0^{\omega_c} d\omega \frac{\chi_{B1g}^{\prime\prime}(\omega,T)}{\omega},
  \label{eq:K-K}
\end{equation}
which projects the low energy part of the spectra but is a well-defined quantity only if $\chi_{B1g}^{\prime\prime}(\omega,T)$ decays for $\omega\to \infty$ and if $\omega_c \to \infty$.

We use $\tilde{R}_{\mu}(T)$ as a normalization factor for extracting scattering rates $\Gamma_\mu(\Omega,T)$ in absolute energy units \cite{Opel:2000}. Since the constant $R$ depends on the experiment the magnitude of $\tilde{R}_{\mu}(T)$ has no direct meaning but compensates for other intensity-dependent quantities when calculating $\Gamma_\mu(\Omega,T)$. A temperature dependent $\tilde{R}_{\mu}(T)$ usually reflects the appearance of an additional scattering channel such as pair breaking or critical fluctuations below $T_c$ or $T_f$, respectively.

\begin{figure}[tbp]
  \centering
  \includegraphics[width=1.0\columnwidth]{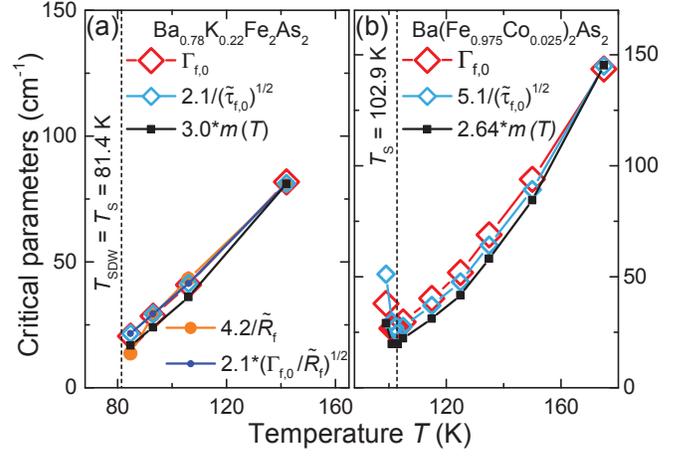}
  \caption{Parameters of the fluctuation response in (a) BKFA and (b) BFCA. ${\Gamma}_{\rm f,0}(T)$ is given in absolute units and determines the scale. The mass $m(T)$ differs from ${\Gamma}_{\rm f,0}(T)$ by 3 and {2.64} for BKFA and BFCA, respectively. The inverse moment $1/\tilde{R_{\rm f}}$ scales as ${\Gamma}_{\rm f,0}(T)$ as does $\left[\tilde{\tau}_{\rm f,0}(T)\right]^{-1/2}$.
  }
  \label{fig:parameters}
\end{figure}

Constant $\tilde{R}_{B1g}(T)$, as observed here in $B_{1g}$ symmetry above the onset of fluctuations at $T_f$, suggests that only one channel contributes to the response. Since the relaxation rate $\Gamma_{B1g}(\Omega\to 0,T)=\Gamma_{B1g,0}(T)$ as derived from the spectra above $T_f$ via the memory function method \cite{Opel:2000} has the same temperature dependence as the resistivity $\varrho(T)$ we conclude that the response originates in e-h excitations. Below $T_f$, $\tilde{R}_{B1g}(T)$ increases signaling the appearance of critical fluctuations, and $\Gamma_{B1g,0}(T)$ decreases faster than $\varrho(T)$. From the isolated fluctuation response  [Fig.~\ref{fig:BKFA_fluct}\,(a)] we derive $\Gamma_{\rm f,0}(T)$ using the memory function method with a normalization $\tilde{R}_{\rm f}(T)$. Somewhat unexpectedly, we find that $1/\tilde{R}_{\rm f}(T)$ and $\Gamma_{\rm f,0}(T)$ have an almost identical temperature dependence (modulo a constant factor) as shown in Fig.~\ref{fig:parameters}\,(a). The nearly linear variation with temperature of both quantities shows that $\tilde{R}_{\rm f}(T)$ and $1/\Gamma_{\rm f,0}(T)$ are critical and approximately proportional to $|T-T_0|^{-1}$.

On the other hand, the initial slope of the fluctuational response,
\begin{equation}
  \tilde{\tau}_{\rm f,0}(T) = \left.R\frac{\partial\chi_{\rm f}^{\prime\prime}(\Omega,T)}{\partial\Omega}\right|_{\Omega=0},
  \label{eq:slope}
\end{equation}
can be extracted by plotting $\lim_{\Omega\to 0}\left[R\chi_{\rm f}^{\prime\prime}(\Omega,T)/\Omega\right]$ as demonstrated by Kretzschmar and coworkers \cite{Kretzschmar:2016}.  $\tilde{\tau}_{\rm f,0}(T)$ is again an $R$-dependent quantity. Fig.~\ref{fig:parameters}\,(a) shows that $\tilde{\tau}_{\rm f,0}(T)$ is identical to $\tilde{R}_{\rm f}(T)/{\Gamma}_{\rm f,0}(T)$. From what we saw before the temperature dependence is that of $[\tilde{R}_{\rm f}(T)]^2\propto |T-T_0|^{-2}$. In principle both $R{\chi}^{\prime\prime}_{\rm f,0}(\Omega,T)$ and $\chi^{\rm el}_{\rm nem,0} (T)$ in Eq.~\eqref{eq:raman} can be critical. However, since the overall temperature dependence may indicate double counting, the interrelation of the two functions is not settled and needs to be worked out in a future study.

Fig.~\ref{fig:parameters}\,(b) shows the parameters for BFCA, $n=0.025$. The overall trends are similar to those for BKFA in panel (a).  The masses $m(T)$ which can be derived in absolute energy units are different from ${\Gamma}_{\rm f,0}(T)$ by factors between 2 and 3 but exhibit {qualitatively similar} temperature dependences {(for the detailed doping dependence see Fig.~\ref{fig:mass})}.

\begin{figure}[tbp]
  \centering
  \includegraphics[width=1.0\columnwidth]{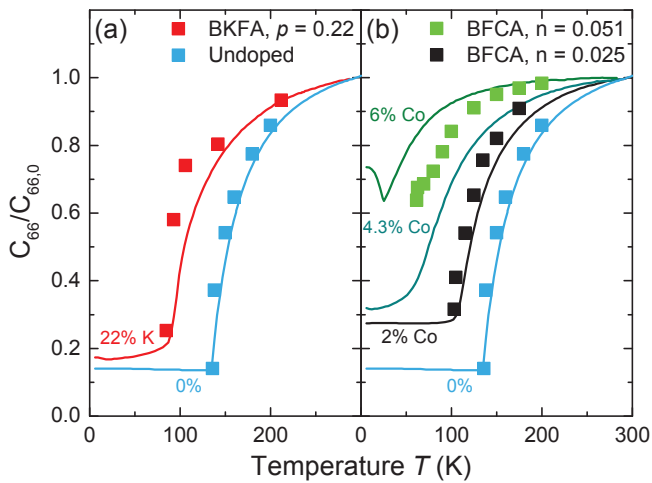}
  \caption{Nematic susceptibility and elastic constants. The temperature dependence of $C_{66}(T)/C_{66,0}$ is taken from Ref. \cite{Bohmer:2016}. The data points are calculated via Eq.~\eqref{eq:C66}.
  }
  \label{fig:nem}
\end{figure}

In Fig.~\ref{fig:nem} we now compare $C_{66}$ with $\chi^{\rm el}_{\rm nem,0}(T)$ according to Eq.~\eqref{eq:C66}. We find that the temperature dependence of the initial slope of the fluctuation response (Eq.~\ref{eq:slope}) is too strong for both BFCA and BKFA. $1/{\Gamma}_{\rm f,0}(T)$, on the other hand, leads to a satisfactory agreement for BFCA as expected because of the proportionality of $1/{\Gamma}_{\rm f,0}(T)$ and $\tilde{R}_{\rm f}(T)$ and thus corroborates the analysis presented in Ref.~\cite{Gallais:2016}. We prefer to use $1/{\Gamma}_{\rm f,0}(T)$, for having absolute units, and hope that the coupling constant $\lambda$ can be derived in the future. In contrast to the results in BFCA and the parent compound the temperature dependences derived for $C_{66}$ from the thermodynamic and the Raman measurements show significant differences in  BKFA [Fig.~\ref{fig:nem}\,(a)]. Although the fluctuation response is weak in BKFA we consider the deviations significant.

Finally, it would be desirable to distinguish between the two mechanisms at the origin of the fluctuation response. Following the work of Caprara \textit{et al.} \cite{Caprara:2005}, the data here were analyzed in terms of the exchange of two critical fluctuations with wave vector $\pm{\bf q}_c$ as described first by Aslamazov and Larkin (AL) \cite{Aslamasov:1968}. The evaluation of the diagrams leads to a contribution to the response independent of that of the e-h excitations. Since intermediate electronic states are involved all types of fluctuations can couple to the light. No assumptions as to the origin of the fluctuations and their wave vector or momentum conservation are necessary. Rather, the selection rules are part of the solution and the $q=0$ limit applies automatically in the case of the exchange of two fluctuations with opposite momenta.

If the conduction electrons couple to a single fluctuation restrictions apply as to the momentum conservation and the separability of the various contributions to the response. Only for fluctuations which do not break the full translational symmetry of the lattice such as for ferro-orbital or quadrupolar charge fluctuations momentum conservation is maintained. In all other cases a momentum of order $|{\bf q}_c|$ needs to be supplied corresponding to a mean free path of the carriers $\ell=v_F\tau$ with $v_F$ the Fermi velocity. The relaxation rate $\tau^{-1}$ may come from sources other than  impurities but this is the only case which has been analyzed so far  \cite{Gallais:2016}. In any case, the response vanishes identically for $\tau^{-1}\to 0$ (collision-less limit). For finite $\tau^{-1}$, the spectral shape is entirely given by relaxation behavior of the carriers at high temperature, $T\gg T_0$. For $T\to T_0$, $\tau^{-1}(T)$ will be renormalized, becomes critical, and vanishes as $|T-T_0|$. The resulting response diverges as $1/\Omega$, and $\tilde{R}_{B1g}(T)\propto |T-T_0|^{-1}$. Eq.~(41) and Fig.~10\,(b) of Ref.~\cite{Gallais:2016} allow an estimate for $\tau^{-1}$ yielding $100<\hbar\tau^{-1}<200\,{\rm cm}^{-1}$ independent of doping. An impurity scattering rate of this magnitude is unrealistic since the pair-breaking feature below $T_c$ would be suppressed proportional to $\Delta\tau$, with $\Delta<\hbar\tau^{-1}$ the energy gap, and would become unobservable \cite{Devereaux:1992,Devereaux:1995} (see also next paragraph).
For dynamical electron scattering, $\tau \to \tau(\Omega,T)$, the momentum can be carried away but there is no detailed theoretical study yet. Independent of whether the scattering is elastic or inelastic the fluctuations and the e-h excitations cannot be disentangled.

Several of these issues disappear if the fluctuations are analyzed in terms of AL diagrams. In particular, as shown in Figs.~\ref{fig:BKFA}\,(d) and \ref{fig:BKFA_fluct}\,(b) both the e-h continuum and the fluctuations can be described with realistic parameters. In particular, the relaxation rate derived for the e-h continuum fits that obtained from the resistivity [Fig.~\ref{fig:BKFA}\,(d)], and the quantities derived from the fluctuations are compatible with $C_{66}$. Although the spin dynamics is closely intertwined with charge fluctuations our analysis of the Raman response in terms of the exchange of two spin fluctuations with momenta $\pm{\bf q}_c$ is supported by several arguments in particular in BFCA and by neutron scattering experiments \cite{Inosov:2016}. On the basis of the presently available data we therefore consider it more likely.

\subsection{{Fluctuations and superconductivity}}

The question as to the influence of fluctuations on Cooper pairing is probably even more tantalizing than that on the phase transitions. Around optimal doping a QCP was proposed to exist above which the fluctuations are particularly strong and can support Cooper pairing \cite{Lederer:2015}. In this case the spectral weight in the $B_{1g}$ pair breaking peak is predicted to increase along with  $T_c$ if the doping decreases from the disordered side towards the QCP \cite{Gallais:2015} whereas the spectral weight in the other symmetries should exhibit little dependence on doping.

\begin{figure}[tbp]
  \centering
  \includegraphics[width=0.8\columnwidth]{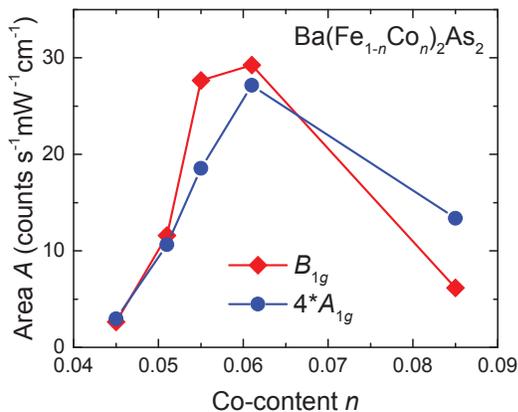}
  \caption{Spectral weight of the pair-breaking maxima in $A_{1g}$ and $B_{1g}$ symmetry as indicated. The spectral weight in $A_{1g}$ symmetry is multiplied by 4.
  }
  \label{fig:BFCA_sw}
\end{figure}

Fig.~\ref{fig:BFCA} displays the pair-breaking effect of BFCA for $0.041\le n \le 0.085$ including optimal doping  at $n \approx 0.06$. In $B_{2g}$ symmetry [Fig.~\ref{fig:BFCA}\,(k)--(o)] we cannot detect any differences between the normal and the superconducting state for reasons discussed earlier \cite{Mazin:2010a}. In both $A_{1g}$ and $B_{1g}$ symmetry a gap and the pair-breaking effect can be observed. For quantifying the spectral weight we integrated the difference between the superconducting and the normal spectra between the intersection point and the high-energy limit of the measurements [grey-shaded areas in Fig.~\ref{fig:BFCA}\,(a)--(j)]. The area is approximately four times larger in $B_{1g}$ than in $A_{1g}$ symmetry. However, as shown in Fig.~\ref{fig:BFCA_sw}, the doping dependences are similar and exhibit maxima at $n=0.061$. What kind of explanation could be compatible with the findings shown in Fig.~\ref{fig:BFCA_sw}?

An enhancement of the $B_{1g}$ spectral weight close to optimal doping  can originate in a contribution from fluctuations to Cooper pairing. Then a nematic resonance close to optimal doping can be expected since the fluctuations are strongest around a putative QCP close to the doping $n$ where the phase transition line $T_{\rm SDW}(n)$ approaches zero. Gallais and coworkers \cite{Gallais:2015} argue that the spectral weight in $B_{1g}$ symmetry peaks at the maximal $T_c$ right above the QCP. Here, we observe that the variation of the spectral weight in $A_{1g}$ symmetry has the same doping dependence as that in $B_{1g}$ symmetry and propose an alternative scenario.

There are two trends in BFCA which can reduce the spectral weight independent of fluctuations: the opening of the SDW gap and impurities. (i) For $n<0.06$ an SDW gap opens up. Since it is most likely the result of Fermi surface nesting it should open up on the electron and the hole bands as suggested by the observation of a redistribution of spectral weight in all symmetries. The opening of the SDW gap explains the reduction of spectral weight on the underdoped side in a natural way since parts of the Fermi surfaces become gapped already above $T_c$. (ii) On the overdoped side the concentration of impurities in the Fe planes becomes substantial and reduces the pair-breaking maxima being proportional to $\Delta\tau$ where $\tau$ is the impurity scattering time \cite{Devereaux:1995,Devereaux:1993}. In addition, a putative accidental node is lifted due to scattering between Fermi surface parts with small and large gaps. Possibly, indications of both effects of impurities are found in BFCA at $n=0.085$ where the pair-breaking peak is reduced and a finite gap appears at low energies \cite{Muschler:2009}. This explanation does not support an interrelation between fluctuations and Cooper pairing but is more compatible with the doping dependence of the spectral weights in $B_{1g}$ and $A_{1g}$ symmetry.

\section{Conclusions}
We have presented light scattering results of BKFA in the normal state and BFCA below $T_c$. In underdoped BKFA ($p=0.22$) we find  the response of critical spin fluctuations similar {but not equal} to that in BFCA. The intensity {in BKFA} is weaker than in BFCA { and the critical exponent $\nu$ in the expression for the mass [see Eq.~\eqref{eq:mass}] depends strongly on doping. The residual mass at $T_s$ decreases with $T_s$ indicating a QCP on either side of zero doping}. The temperature dependence {of the fluctuation response is}, by and large, consistent with the variation of the shear modulus $C_{66}(T)$. The problem arises which of the derived quantities, the first moment of the response, $\tilde{R}_{f}(T)$, the static relaxation rate, ${\Gamma}_{\rm f,0}(T)$, or the initial slope of the fluctuation response, $\tilde{\tau}_{\rm f,0}(T)$, should be used for the comparison with the thermodynamical data. We find that ${\Gamma}_{\rm f,0}(T)$ has the same temperature dependence as $\tilde{R}_{f}(T)$ and describes $C_{66}(T)$ best in addition to being available in absolute energy units. It is not clear at the moment as to why $\tilde{\tau}_{\rm f,0}(T) \propto {\Gamma}^{-2}_{\rm f,0}(T)$ is inappropriate for describing $C_{66}(T)$. Possibly the interaction between the fluctuations leads to an additional factor $|T-T_0|^{-1}$ which appears only in the Raman response. More work is needed here.

In the superconducting state of BFCA we find a strong but similar doping dependence of the spectral weights of the pair-breaking maxima of both the $A_{1g}$ and the $B_{1g}$ spectra peaking at $n\approx 0.06$. Therefore, the $B_{1g}$ response is not specifically enhanced as one would expect for an interrelation of  nematic fluctuations and Cooper pairing \cite{Gallais:2015}. One has to conclude that signatures of unconventional pairing channels in experimental probes remain rare and indirect such as the observation of a spin resonance \cite{Inosov:2016,Christianson:2008} and of a Bardasis-Schrieffer mode in optimally doped BKFA \cite{Bohm:2014} and possibly BFCA indicating substantial attraction in the subleading $d_{x^2-y^2}$ channel on top of the $s_\pm$ (or $s_{++}$) ground state.

\begin{acknowledgements}
We acknowledge useful discussions with D. Einzel, L. Benfatto, C. Meingast, and J. Schmalian. Financial support for the work came from the DFG via the Priority Program SPP\,1458 (project no. HA\,2071/7, the Bavarian Californian Technology Center BaCaTeC (project no. A5\,[2012-2]), and from the Transregional Collaborative Research Center TRR\,80. Work in the SIMES at Stanford University and SLAC was supported by the U.S. Department of Energy, Office of Basic Energy Sciences, Division of Materials Sciences and Engineering, under Contract No. DE-AC02-76SF00515.
\end{acknowledgements}


\begin{thebibliography}{60}%
\makeatletter
\providecommand \@ifxundefined [1]{%
 \@ifx{#1\undefined}
}%
\providecommand \@ifnum [1]{%
 \ifnum #1\expandafter \@firstoftwo
 \else \expandafter \@secondoftwo
 \fi
}%
\providecommand \@ifx [1]{%
 \ifx #1\expandafter \@firstoftwo
 \else \expandafter \@secondoftwo
 \fi
}%
\providecommand \natexlab [1]{#1}%
\providecommand \enquote  [1]{``#1''}%
\providecommand \bibnamefont  [1]{#1}%
\providecommand \bibfnamefont [1]{#1}%
\providecommand \citenamefont [1]{#1}%
\providecommand \href@noop [0]{\@secondoftwo}%
\providecommand \href [0]{\begingroup \@sanitize@url \@href}%
\providecommand \@href[1]{\@@startlink{#1}\@@href}%
\providecommand \@@href[1]{\endgroup#1\@@endlink}%
\providecommand \@sanitize@url [0]{\catcode `\\12\catcode `\$12\catcode
  `\&12\catcode `\#12\catcode `\^12\catcode `\_12\catcode `\%12\relax}%
\providecommand \@@startlink[1]{}%
\providecommand \@@endlink[0]{}%
\providecommand \url  [0]{\begingroup\@sanitize@url \@url }%
\providecommand \@url [1]{\endgroup\@href {#1}{\urlprefix }}%
\providecommand \urlprefix  [0]{URL }%
\providecommand \Eprint [0]{\href }%
\providecommand \doibase [0]{http://dx.doi.org/}%
\providecommand \selectlanguage [0]{\@gobble}%
\providecommand \bibinfo  [0]{\@secondoftwo}%
\providecommand \bibfield  [0]{\@secondoftwo}%
\providecommand \translation [1]{[#1]}%
\providecommand \BibitemOpen [0]{}%
\providecommand \bibitemStop [0]{}%
\providecommand \bibitemNoStop [0]{.\EOS\space}%
\providecommand \EOS [0]{\spacefactor3000\relax}%
\providecommand \BibitemShut  [1]{\csname bibitem#1\endcsname}%
\let\auto@bib@innerbib\@empty
\bibitem [{\citenamefont {Suhl}\ \emph {et~al.}(1959)\citenamefont {Suhl},
  \citenamefont {Matthias},\ and\ \citenamefont {Walker}}]{Suhl:1959}%
  \BibitemOpen
  \bibfield  {author} {\bibinfo {author} {\bibfnamefont {H.}~\bibnamefont
  {Suhl}}, \bibinfo {author} {\bibfnamefont {B.~T.}\ \bibnamefont {Matthias}},
  \ and\ \bibinfo {author} {\bibfnamefont {L.~R.}\ \bibnamefont {Walker}},\
  }\href {\doibase 10.1103/PhysRevLett.3.552} {\bibfield  {journal} {\bibinfo
  {journal} {Phys. Rev. Lett.}\ }\textbf {\bibinfo {volume} {3}},\ \bibinfo
  {pages} {552} (\bibinfo {year} {1959})}\BibitemShut {NoStop}%
\bibitem [{\citenamefont {Carbotte}(1990)}]{Carbotte:1990}%
  \BibitemOpen
  \bibfield  {author} {\bibinfo {author} {\bibfnamefont {J.~P.}\ \bibnamefont
  {Carbotte}},\ }\href {\doibase 10.1103/RevModPhys.62.1027} {\bibfield
  {journal} {\bibinfo  {journal} {Rev. Mod. Phys.}\ }\textbf {\bibinfo {volume}
  {62}},\ \bibinfo {pages} {1027} (\bibinfo {year} {1990})}\BibitemShut
  {NoStop}%
\bibitem [{\citenamefont {Johnston}\ \emph {et~al.}(2010)\citenamefont
  {Johnston}, \citenamefont {Vernay}, \citenamefont {Moritz}, \citenamefont
  {Shen}, \citenamefont {Nagaosa}, \citenamefont {Zaanen},\ and\ \citenamefont
  {Devereaux}}]{Johnston:2010a}%
  \BibitemOpen
  \bibfield  {author} {\bibinfo {author} {\bibfnamefont {S.}~\bibnamefont
  {Johnston}}, \bibinfo {author} {\bibfnamefont {F.}~\bibnamefont {Vernay}},
  \bibinfo {author} {\bibfnamefont {B.}~\bibnamefont {Moritz}}, \bibinfo
  {author} {\bibfnamefont {Z.-X.}\ \bibnamefont {Shen}}, \bibinfo {author}
  {\bibfnamefont {N.}~\bibnamefont {Nagaosa}}, \bibinfo {author} {\bibfnamefont
  {J.}~\bibnamefont {Zaanen}}, \ and\ \bibinfo {author} {\bibfnamefont {T.~P.}\
  \bibnamefont {Devereaux}},\ }\href {\doibase 10.1103/PhysRevB.82.064513}
  {\bibfield  {journal} {\bibinfo  {journal} {Phys. Rev. B}\ }\textbf {\bibinfo
  {volume} {82}},\ \bibinfo {pages} {064513} (\bibinfo {year}
  {2010})}\BibitemShut {NoStop}%
\bibitem [{\citenamefont {Ge}\ \emph {et~al.}(2015)\citenamefont {Ge},
  \citenamefont {Liu}, \citenamefont {Liu}, \citenamefont {Gao}, \citenamefont
  {Qian}, \citenamefont {Xue}, \citenamefont {Liu},\ and\ \citenamefont
  {Jia}}]{GeJF:2015}%
  \BibitemOpen
  \bibfield  {author} {\bibinfo {author} {\bibfnamefont {J.-F.}\ \bibnamefont
  {Ge}}, \bibinfo {author} {\bibfnamefont {Z.-L.}\ \bibnamefont {Liu}},
  \bibinfo {author} {\bibfnamefont {C.}~\bibnamefont {Liu}}, \bibinfo {author}
  {\bibfnamefont {C.-L.}\ \bibnamefont {Gao}}, \bibinfo {author} {\bibfnamefont
  {D.}~\bibnamefont {Qian}}, \bibinfo {author} {\bibfnamefont {Q.-K.}\
  \bibnamefont {Xue}}, \bibinfo {author} {\bibfnamefont {Y.}~\bibnamefont
  {Liu}}, \ and\ \bibinfo {author} {\bibfnamefont {J.-F.}\ \bibnamefont
  {Jia}},\ }\href {\doibase 10.1038/nmat4153} {\bibfield  {journal} {\bibinfo
  {journal} {Nature Mater.}\ }\textbf {\bibinfo {volume} {14}},\ \bibinfo
  {pages} {285} (\bibinfo {year} {2015})}\BibitemShut {NoStop}%
\bibitem [{\citenamefont {{Lee}}\ \emph {et~al.}(2014)\citenamefont {{Lee}},
  \citenamefont {{Schmitt}}, \citenamefont {{Moore}}, \citenamefont
  {{Johnston}}, \citenamefont {{Cui}}, \citenamefont {{Li}}, \citenamefont
  {{Yi}}, \citenamefont {{Liu}}, \citenamefont {{Hashimoto}}, \citenamefont
  {{Zhang}}, \citenamefont {{Lu}}, \citenamefont {{Devereaux}}, \citenamefont
  {{Lee}},\ and\ \citenamefont {{Shen}}}]{LeeJJ:2014}%
  \BibitemOpen
  \bibfield  {author} {\bibinfo {author} {\bibfnamefont {J.~J.}\ \bibnamefont
  {{Lee}}}, \bibinfo {author} {\bibfnamefont {F.~T.}\ \bibnamefont
  {{Schmitt}}}, \bibinfo {author} {\bibfnamefont {R.~G.}\ \bibnamefont
  {{Moore}}}, \bibinfo {author} {\bibfnamefont {S.}~\bibnamefont {{Johnston}}},
  \bibinfo {author} {\bibfnamefont {Y.-T.}\ \bibnamefont {{Cui}}}, \bibinfo
  {author} {\bibfnamefont {W.}~\bibnamefont {{Li}}}, \bibinfo {author}
  {\bibfnamefont {M.}~\bibnamefont {{Yi}}}, \bibinfo {author} {\bibfnamefont
  {Z.~K.}\ \bibnamefont {{Liu}}}, \bibinfo {author} {\bibfnamefont
  {M.}~\bibnamefont {{Hashimoto}}}, \bibinfo {author} {\bibfnamefont
  {Y.}~\bibnamefont {{Zhang}}}, \bibinfo {author} {\bibfnamefont {D.~H.}\
  \bibnamefont {{Lu}}}, \bibinfo {author} {\bibfnamefont {T.~P.}\ \bibnamefont
  {{Devereaux}}}, \bibinfo {author} {\bibfnamefont {D.-H.}\ \bibnamefont
  {{Lee}}}, \ and\ \bibinfo {author} {\bibfnamefont {Z.-X.}\ \bibnamefont
  {{Shen}}},\ }\href {\doibase 10.1038/nature13894} {\bibfield  {journal}
  {\bibinfo  {journal} {Nature}\ }\textbf {\bibinfo {volume} {515}},\ \bibinfo
  {pages} {245} (\bibinfo {year} {2014})}\BibitemShut {NoStop}%
\bibitem [{\citenamefont {Lederer}\ \emph {et~al.}(2015)\citenamefont
  {Lederer}, \citenamefont {Schattner}, \citenamefont {Berg},\ and\
  \citenamefont {Kivelson}}]{Lederer:2015}%
  \BibitemOpen
  \bibfield  {author} {\bibinfo {author} {\bibfnamefont {S.}~\bibnamefont
  {Lederer}}, \bibinfo {author} {\bibfnamefont {Y.}~\bibnamefont {Schattner}},
  \bibinfo {author} {\bibfnamefont {E.}~\bibnamefont {Berg}}, \ and\ \bibinfo
  {author} {\bibfnamefont {S.~A.}\ \bibnamefont {Kivelson}},\ }\href {\doibase
  10.1103/PhysRevLett.114.097001} {\bibfield  {journal} {\bibinfo  {journal}
  {Phys. Rev. Lett.}\ }\textbf {\bibinfo {volume} {114}},\ \bibinfo {pages}
  {097001} (\bibinfo {year} {2015})}\BibitemShut {NoStop}%
\bibitem [{\citenamefont {Muschler}\ \emph {et~al.}(2010)\citenamefont
  {Muschler}, \citenamefont {Prestel}, \citenamefont {Schachinger},
  \citenamefont {Carbotte}, \citenamefont {Hackl}, \citenamefont {Ono},\ and\
  \citenamefont {Ando}}]{Muschler:2010}%
  \BibitemOpen
  \bibfield  {author} {\bibinfo {author} {\bibfnamefont {B.}~\bibnamefont
  {Muschler}}, \bibinfo {author} {\bibfnamefont {W.}~\bibnamefont {Prestel}},
  \bibinfo {author} {\bibfnamefont {E.}~\bibnamefont {Schachinger}}, \bibinfo
  {author} {\bibfnamefont {J.~P.}\ \bibnamefont {Carbotte}}, \bibinfo {author}
  {\bibfnamefont {R.}~\bibnamefont {Hackl}}, \bibinfo {author} {\bibfnamefont
  {S.}~\bibnamefont {Ono}}, \ and\ \bibinfo {author} {\bibfnamefont
  {Y.}~\bibnamefont {Ando}},\ }\href {\doibase 10.1088/0953-8984/22/37/375702}
  {\bibfield  {journal} {\bibinfo  {journal} {J. Phys. Condens. Matter}\
  }\textbf {\bibinfo {volume} {22}},\ \bibinfo {pages} {375702} (\bibinfo
  {year} {2010})}\BibitemShut {NoStop}%
\bibitem [{\citenamefont {Caprara}\ \emph {et~al.}(2011)\citenamefont
  {Caprara}, \citenamefont {Di~Castro}, \citenamefont {Muschler}, \citenamefont
  {Prestel}, \citenamefont {Hackl}, \citenamefont {Lambacher}, \citenamefont
  {Erb}, \citenamefont {Komiya}, \citenamefont {Ando},\ and\ \citenamefont
  {Grilli}}]{Caprara:2011}%
  \BibitemOpen
  \bibfield  {author} {\bibinfo {author} {\bibfnamefont {S.}~\bibnamefont
  {Caprara}}, \bibinfo {author} {\bibfnamefont {C.}~\bibnamefont {Di~Castro}},
  \bibinfo {author} {\bibfnamefont {B.}~\bibnamefont {Muschler}}, \bibinfo
  {author} {\bibfnamefont {W.}~\bibnamefont {Prestel}}, \bibinfo {author}
  {\bibfnamefont {R.}~\bibnamefont {Hackl}}, \bibinfo {author} {\bibfnamefont
  {M.}~\bibnamefont {Lambacher}}, \bibinfo {author} {\bibfnamefont
  {A.}~\bibnamefont {Erb}}, \bibinfo {author} {\bibfnamefont {S.}~\bibnamefont
  {Komiya}}, \bibinfo {author} {\bibfnamefont {Y.}~\bibnamefont {Ando}}, \ and\
  \bibinfo {author} {\bibfnamefont {M.}~\bibnamefont {Grilli}},\ }\href
  {\doibase 10.1103/PhysRevB.84.054508} {\bibfield  {journal} {\bibinfo
  {journal} {Phys. Rev. B}\ }\textbf {\bibinfo {volume} {84}},\ \bibinfo
  {pages} {054508} (\bibinfo {year} {2011})}\BibitemShut {NoStop}%
\bibitem [{\citenamefont {Kretzschmar}\ \emph {et~al.}(2013)\citenamefont
  {Kretzschmar}, \citenamefont {Muschler}, \citenamefont {B\"ohm},
  \citenamefont {Baum}, \citenamefont {Hackl}, \citenamefont {Wen},
  \citenamefont {Tsurkan}, \citenamefont {Deisenhofer},\ and\ \citenamefont
  {Loidl}}]{Kretzschmar:2013}%
  \BibitemOpen
  \bibfield  {author} {\bibinfo {author} {\bibfnamefont {F.}~\bibnamefont
  {Kretzschmar}}, \bibinfo {author} {\bibfnamefont {B.}~\bibnamefont
  {Muschler}}, \bibinfo {author} {\bibfnamefont {T.}~\bibnamefont {B\"ohm}},
  \bibinfo {author} {\bibfnamefont {A.}~\bibnamefont {Baum}}, \bibinfo {author}
  {\bibfnamefont {R.}~\bibnamefont {Hackl}}, \bibinfo {author} {\bibfnamefont
  {H.-H.}\ \bibnamefont {Wen}}, \bibinfo {author} {\bibfnamefont
  {V.}~\bibnamefont {Tsurkan}}, \bibinfo {author} {\bibfnamefont
  {J.}~\bibnamefont {Deisenhofer}}, \ and\ \bibinfo {author} {\bibfnamefont
  {A.}~\bibnamefont {Loidl}},\ }\href {\doibase 10.1103/PhysRevLett.110.187002}
  {\bibfield  {journal} {\bibinfo  {journal} {Phys. Rev. Lett.}\ }\textbf
  {\bibinfo {volume} {110}},\ \bibinfo {pages} {187002} (\bibinfo {year}
  {2013})}\BibitemShut {NoStop}%
\bibitem [{\citenamefont {B\"ohm}\ \emph {et~al.}(2014)\citenamefont {B\"ohm},
  \citenamefont {Kemper}, \citenamefont {Moritz}, \citenamefont {Kretzschmar},
  \citenamefont {Muschler}, \citenamefont {Eiter}, \citenamefont {Hackl},
  \citenamefont {Devereaux}, \citenamefont {Scalapino},\ and\ \citenamefont
  {Wen}}]{Bohm:2014}%
  \BibitemOpen
  \bibfield  {author} {\bibinfo {author} {\bibfnamefont {T.}~\bibnamefont
  {B\"ohm}}, \bibinfo {author} {\bibfnamefont {A.~F.}\ \bibnamefont {Kemper}},
  \bibinfo {author} {\bibfnamefont {B.}~\bibnamefont {Moritz}}, \bibinfo
  {author} {\bibfnamefont {F.}~\bibnamefont {Kretzschmar}}, \bibinfo {author}
  {\bibfnamefont {B.}~\bibnamefont {Muschler}}, \bibinfo {author}
  {\bibfnamefont {H.-M.}\ \bibnamefont {Eiter}}, \bibinfo {author}
  {\bibfnamefont {R.}~\bibnamefont {Hackl}}, \bibinfo {author} {\bibfnamefont
  {T.~P.}\ \bibnamefont {Devereaux}}, \bibinfo {author} {\bibfnamefont {D.~J.}\
  \bibnamefont {Scalapino}}, \ and\ \bibinfo {author} {\bibfnamefont {H.-H.}\
  \bibnamefont {Wen}},\ }\href {\doibase 10.1103/PhysRevX.4.041046} {\bibfield
  {journal} {\bibinfo  {journal} {Phys. Rev. X}\ }\textbf {\bibinfo {volume}
  {4}},\ \bibinfo {pages} {041046} (\bibinfo {year} {2014})}\BibitemShut
  {NoStop}%
\bibitem [{\citenamefont {Caprara}\ \emph {et~al.}(2005)\citenamefont
  {Caprara}, \citenamefont {Di~Castro}, \citenamefont {Grilli},\ and\
  \citenamefont {Suppa}}]{Caprara:2005}%
  \BibitemOpen
  \bibfield  {author} {\bibinfo {author} {\bibfnamefont {S.}~\bibnamefont
  {Caprara}}, \bibinfo {author} {\bibfnamefont {C.}~\bibnamefont {Di~Castro}},
  \bibinfo {author} {\bibfnamefont {M.}~\bibnamefont {Grilli}}, \ and\ \bibinfo
  {author} {\bibfnamefont {D.}~\bibnamefont {Suppa}},\ }\href {\doibase
  10.1103/PhysRevLett.95.117004} {\bibfield  {journal} {\bibinfo  {journal}
  {Phys. Rev. Lett.}\ }\textbf {\bibinfo {volume} {95}},\ \bibinfo {pages}
  {117004} (\bibinfo {year} {2005})}\BibitemShut {NoStop}%
\bibitem [{\citenamefont {Tassini}\ \emph {et~al.}(2005)\citenamefont
  {Tassini}, \citenamefont {Venturini}, \citenamefont {Zhang}, \citenamefont
  {Hackl}, \citenamefont {Kikugawa},\ and\ \citenamefont
  {Fujita}}]{Tassini:2005}%
  \BibitemOpen
  \bibfield  {author} {\bibinfo {author} {\bibfnamefont {L.}~\bibnamefont
  {Tassini}}, \bibinfo {author} {\bibfnamefont {F.}~\bibnamefont {Venturini}},
  \bibinfo {author} {\bibfnamefont {Q.-M.}\ \bibnamefont {Zhang}}, \bibinfo
  {author} {\bibfnamefont {R.}~\bibnamefont {Hackl}}, \bibinfo {author}
  {\bibfnamefont {N.}~\bibnamefont {Kikugawa}}, \ and\ \bibinfo {author}
  {\bibfnamefont {T.}~\bibnamefont {Fujita}},\ }\href {\doibase
  10.1103/PhysRevLett.95.117002} {\bibfield  {journal} {\bibinfo  {journal}
  {Phys. Rev. Lett.}\ }\textbf {\bibinfo {volume} {95}},\ \bibinfo {pages}
  {117002} (\bibinfo {year} {2005})}\BibitemShut {NoStop}%
\bibitem [{\citenamefont {Eiter}\ \emph {et~al.}(2013)\citenamefont {Eiter},
  \citenamefont {Lavagnini}, \citenamefont {Hackl}, \citenamefont {Nowadnick},
  \citenamefont {Kemper}, \citenamefont {Devereaux}, \citenamefont {Chu},
  \citenamefont {Analytis}, \citenamefont {Fisher},\ and\ \citenamefont
  {Degiorgi}}]{Eiter:2013}%
  \BibitemOpen
  \bibfield  {author} {\bibinfo {author} {\bibfnamefont {H.-M.}\ \bibnamefont
  {Eiter}}, \bibinfo {author} {\bibfnamefont {M.}~\bibnamefont {Lavagnini}},
  \bibinfo {author} {\bibfnamefont {R.}~\bibnamefont {Hackl}}, \bibinfo
  {author} {\bibfnamefont {E.~A.}\ \bibnamefont {Nowadnick}}, \bibinfo {author}
  {\bibfnamefont {A.~F.}\ \bibnamefont {Kemper}}, \bibinfo {author}
  {\bibfnamefont {T.~P.}\ \bibnamefont {Devereaux}}, \bibinfo {author}
  {\bibfnamefont {J.-H.}\ \bibnamefont {Chu}}, \bibinfo {author} {\bibfnamefont
  {J.~G.}\ \bibnamefont {Analytis}}, \bibinfo {author} {\bibfnamefont {I.~R.}\
  \bibnamefont {Fisher}}, \ and\ \bibinfo {author} {\bibfnamefont
  {L.}~\bibnamefont {Degiorgi}},\ }\href {\doibase 10.1073/pnas.1214745110}
  {\bibfield  {journal} {\bibinfo  {journal} {Proc. Nat. Acad. Sciences}\
  }\textbf {\bibinfo {volume} {110}},\ \bibinfo {pages} {64} (\bibinfo {year}
  {2013})},\ \Eprint
  {http://arxiv.org/abs/http://www.pnas.org/content/110/1/64.full.pdf+html}
  {http://www.pnas.org/content/110/1/64.full.pdf+html} \BibitemShut {NoStop}%
\bibitem [{\citenamefont {Perali}\ \emph {et~al.}(1996)\citenamefont {Perali},
  \citenamefont {Castellani}, \citenamefont {Di~Castro},\ and\ \citenamefont
  {Grilli}}]{Perali:1996}%
  \BibitemOpen
  \bibfield  {author} {\bibinfo {author} {\bibfnamefont {A.}~\bibnamefont
  {Perali}}, \bibinfo {author} {\bibfnamefont {C.}~\bibnamefont {Castellani}},
  \bibinfo {author} {\bibfnamefont {C.}~\bibnamefont {Di~Castro}}, \ and\
  \bibinfo {author} {\bibfnamefont {M.}~\bibnamefont {Grilli}},\ }\href
  {\doibase 10.1103/PhysRevB.54.16216} {\bibfield  {journal} {\bibinfo
  {journal} {Phys. Rev. B}\ }\textbf {\bibinfo {volume} {54}},\ \bibinfo
  {pages} {16216} (\bibinfo {year} {1996})}\BibitemShut {NoStop}%
\bibitem [{\citenamefont {Gallais}\ \emph {et~al.}(2016)\citenamefont
  {Gallais}, \citenamefont {Paul}, \citenamefont {Chauvi\`ere},\ and\
  \citenamefont {Schmalian}}]{Gallais:2015}%
  \BibitemOpen
  \bibfield  {author} {\bibinfo {author} {\bibfnamefont {Y.}~\bibnamefont
  {Gallais}}, \bibinfo {author} {\bibfnamefont {I.}~\bibnamefont {Paul}},
  \bibinfo {author} {\bibfnamefont {L.}~\bibnamefont {Chauvi\`ere}}, \ and\
  \bibinfo {author} {\bibfnamefont {J.}~\bibnamefont {Schmalian}},\ }\href
  {\doibase 10.1103/PhysRevLett.116.017001} {\bibfield  {journal} {\bibinfo
  {journal} {Phys. Rev. Lett.}\ }\textbf {\bibinfo {volume} {116}},\ \bibinfo
  {pages} {017001} (\bibinfo {year} {2016})}\BibitemShut {NoStop}%
\bibitem [{\citenamefont {{Kretzschmar}}\ \emph {et~al.}(2016)\citenamefont
  {{Kretzschmar}}, \citenamefont {{B{\"o}hm}}, \citenamefont
  {{Karahasanovi{\'c}}}, \citenamefont {{Muschler}}, \citenamefont {{Baum}},
  \citenamefont {{Jost}}, \citenamefont {{Schmalian}}, \citenamefont
  {{Caprara}}, \citenamefont {{Grilli}}, \citenamefont {{Di Castro}},
  \citenamefont {{Analytis}}, \citenamefont {{Chu}}, \citenamefont {{Fisher}},\
  and\ \citenamefont {{Hackl}}}]{Kretzschmar:2016}%
  \BibitemOpen
  \bibfield  {author} {\bibinfo {author} {\bibfnamefont {F.}~\bibnamefont
  {{Kretzschmar}}}, \bibinfo {author} {\bibfnamefont {T.}~\bibnamefont
  {{B{\"o}hm}}}, \bibinfo {author} {\bibfnamefont {U.}~\bibnamefont
  {{Karahasanovi{\'c}}}}, \bibinfo {author} {\bibfnamefont {B.}~\bibnamefont
  {{Muschler}}}, \bibinfo {author} {\bibfnamefont {A.}~\bibnamefont {{Baum}}},
  \bibinfo {author} {\bibfnamefont {D.}~\bibnamefont {{Jost}}}, \bibinfo
  {author} {\bibfnamefont {J.}~\bibnamefont {{Schmalian}}}, \bibinfo {author}
  {\bibfnamefont {S.}~\bibnamefont {{Caprara}}}, \bibinfo {author}
  {\bibfnamefont {M.}~\bibnamefont {{Grilli}}}, \bibinfo {author}
  {\bibfnamefont {C.}~\bibnamefont {{Di Castro}}}, \bibinfo {author}
  {\bibfnamefont {J.~H.}\ \bibnamefont {{Analytis}}}, \bibinfo {author}
  {\bibfnamefont {J.-H.}\ \bibnamefont {{Chu}}}, \bibinfo {author}
  {\bibfnamefont {I.~R.}\ \bibnamefont {{Fisher}}}, \ and\ \bibinfo {author}
  {\bibfnamefont {R.}~\bibnamefont {{Hackl}}},\ }\href {\doibase
  10.1038/NPHYS3634} {\bibfield  {journal} {\bibinfo  {journal} {Nature Phys.}\
  }\textbf {\bibinfo {volume} {12}},\ \bibinfo {pages} {560} (\bibinfo {year}
  {2016})}\BibitemShut {NoStop}%
\bibitem [{\citenamefont {Devereaux}\ \emph {et~al.}(1994)\citenamefont
  {Devereaux}, \citenamefont {Einzel}, \citenamefont {Stadlober}, \citenamefont
  {Hackl}, \citenamefont {Leach},\ and\ \citenamefont
  {Neumeier}}]{Devereaux:1994}%
  \BibitemOpen
  \bibfield  {author} {\bibinfo {author} {\bibfnamefont {T.~P.}\ \bibnamefont
  {Devereaux}}, \bibinfo {author} {\bibfnamefont {D.}~\bibnamefont {Einzel}},
  \bibinfo {author} {\bibfnamefont {B.}~\bibnamefont {Stadlober}}, \bibinfo
  {author} {\bibfnamefont {R.}~\bibnamefont {Hackl}}, \bibinfo {author}
  {\bibfnamefont {D.~H.}\ \bibnamefont {Leach}}, \ and\ \bibinfo {author}
  {\bibfnamefont {J.~J.}\ \bibnamefont {Neumeier}},\ }\href {\doibase
  10.1103/PhysRevLett.72.396} {\bibfield  {journal} {\bibinfo  {journal} {Phys.
  Rev. Lett.}\ }\textbf {\bibinfo {volume} {72}},\ \bibinfo {pages} {396}
  (\bibinfo {year} {1994})}\BibitemShut {NoStop}%
\bibitem [{\citenamefont {Mazin}\ \emph {et~al.}(2010)\citenamefont {Mazin},
  \citenamefont {Devereaux}, \citenamefont {Analytis}, \citenamefont {Chu},
  \citenamefont {Fisher}, \citenamefont {Muschler},\ and\ \citenamefont
  {Hackl}}]{Mazin:2010a}%
  \BibitemOpen
  \bibfield  {author} {\bibinfo {author} {\bibfnamefont {I.~I.}\ \bibnamefont
  {Mazin}}, \bibinfo {author} {\bibfnamefont {T.~P.}\ \bibnamefont
  {Devereaux}}, \bibinfo {author} {\bibfnamefont {J.~G.}\ \bibnamefont
  {Analytis}}, \bibinfo {author} {\bibfnamefont {J.-H.}\ \bibnamefont {Chu}},
  \bibinfo {author} {\bibfnamefont {I.~R.}\ \bibnamefont {Fisher}}, \bibinfo
  {author} {\bibfnamefont {B.}~\bibnamefont {Muschler}}, \ and\ \bibinfo
  {author} {\bibfnamefont {R.}~\bibnamefont {Hackl}},\ }\href {\doibase
  10.1103/PhysRevB.82.180502} {\bibfield  {journal} {\bibinfo  {journal} {Phys.
  Rev. B}\ }\textbf {\bibinfo {volume} {82}},\ \bibinfo {pages} {180502}
  (\bibinfo {year} {2010})}\BibitemShut {NoStop}%
\bibitem [{\citenamefont {Muschler}\ \emph {et~al.}(2009)\citenamefont
  {Muschler}, \citenamefont {Prestel}, \citenamefont {Hackl}, \citenamefont
  {Devereaux}, \citenamefont {Analytis}, \citenamefont {Chu},\ and\
  \citenamefont {Fisher}}]{Muschler:2009}%
  \BibitemOpen
  \bibfield  {author} {\bibinfo {author} {\bibfnamefont {B.}~\bibnamefont
  {Muschler}}, \bibinfo {author} {\bibfnamefont {W.}~\bibnamefont {Prestel}},
  \bibinfo {author} {\bibfnamefont {R.}~\bibnamefont {Hackl}}, \bibinfo
  {author} {\bibfnamefont {T.~P.}\ \bibnamefont {Devereaux}}, \bibinfo {author}
  {\bibfnamefont {J.~G.}\ \bibnamefont {Analytis}}, \bibinfo {author}
  {\bibfnamefont {J.-H.}\ \bibnamefont {Chu}}, \ and\ \bibinfo {author}
  {\bibfnamefont {I.~R.}\ \bibnamefont {Fisher}},\ }\href {\doibase
  10.1103/PhysRevB.80.180510} {\bibfield  {journal} {\bibinfo  {journal} {Phys.
  Rev. B}\ }\textbf {\bibinfo {volume} {80}},\ \bibinfo {pages} {180510}
  (\bibinfo {year} {2009})}\BibitemShut {NoStop}%
\bibitem [{\citenamefont {Chauvi\`ere}\ \emph {et~al.}(2010)\citenamefont
  {Chauvi\`ere}, \citenamefont {Gallais}, \citenamefont {Cazayous},
  \citenamefont {M\'easson}, \citenamefont {Sacuto}, \citenamefont {Colson},\
  and\ \citenamefont {Forget}}]{Chauviere:2010}%
  \BibitemOpen
  \bibfield  {author} {\bibinfo {author} {\bibfnamefont {L.}~\bibnamefont
  {Chauvi\`ere}}, \bibinfo {author} {\bibfnamefont {Y.}~\bibnamefont
  {Gallais}}, \bibinfo {author} {\bibfnamefont {M.}~\bibnamefont {Cazayous}},
  \bibinfo {author} {\bibfnamefont {M.~A.}\ \bibnamefont {M\'easson}}, \bibinfo
  {author} {\bibfnamefont {A.}~\bibnamefont {Sacuto}}, \bibinfo {author}
  {\bibfnamefont {D.}~\bibnamefont {Colson}}, \ and\ \bibinfo {author}
  {\bibfnamefont {A.}~\bibnamefont {Forget}},\ }\href {\doibase
  10.1103/PhysRevB.82.180521} {\bibfield  {journal} {\bibinfo  {journal} {Phys.
  Rev. B}\ }\textbf {\bibinfo {volume} {82}},\ \bibinfo {pages} {180521}
  (\bibinfo {year} {2010})}\BibitemShut {NoStop}%
\bibitem [{\citenamefont {Thorsm\o{}lle}\ \emph {et~al.}(2016)\citenamefont
  {Thorsm\o{}lle}, \citenamefont {Khodas}, \citenamefont {Yin}, \citenamefont
  {Zhang}, \citenamefont {Carr}, \citenamefont {Dai},\ and\ \citenamefont
  {Blumberg}}]{Thorsmolle:2016}%
  \BibitemOpen
  \bibfield  {author} {\bibinfo {author} {\bibfnamefont {V.~K.}\ \bibnamefont
  {Thorsm\o{}lle}}, \bibinfo {author} {\bibfnamefont {M.}~\bibnamefont
  {Khodas}}, \bibinfo {author} {\bibfnamefont {Z.~P.}\ \bibnamefont {Yin}},
  \bibinfo {author} {\bibfnamefont {C.}~\bibnamefont {Zhang}}, \bibinfo
  {author} {\bibfnamefont {S.~V.}\ \bibnamefont {Carr}}, \bibinfo {author}
  {\bibfnamefont {P.}~\bibnamefont {Dai}}, \ and\ \bibinfo {author}
  {\bibfnamefont {G.}~\bibnamefont {Blumberg}},\ }\href {\doibase
  10.1103/PhysRevB.93.054515} {\bibfield  {journal} {\bibinfo  {journal} {Phys.
  Rev. B}\ }\textbf {\bibinfo {volume} {93}},\ \bibinfo {pages} {054515}
  (\bibinfo {year} {2016})}\BibitemShut {NoStop}%
\bibitem [{\citenamefont {Evtushinsky}\ \emph {et~al.}(2009)\citenamefont
  {Evtushinsky}, \citenamefont {Inosov}, \citenamefont {Zabolotnyy},
  \citenamefont {Koitzsch}, \citenamefont {Knupfer}, \citenamefont {B\"uchner},
  \citenamefont {Viazovska}, \citenamefont {Sun}, \citenamefont {Hinkov},
  \citenamefont {Boris}, \citenamefont {Lin}, \citenamefont {Keimer},
  \citenamefont {Varykhalov}, \citenamefont {Kordyuk},\ and\ \citenamefont
  {Borisenko}}]{Evtushinsky:2009}%
  \BibitemOpen
  \bibfield  {author} {\bibinfo {author} {\bibfnamefont {D.~V.}\ \bibnamefont
  {Evtushinsky}}, \bibinfo {author} {\bibfnamefont {D.~S.}\ \bibnamefont
  {Inosov}}, \bibinfo {author} {\bibfnamefont {V.~B.}\ \bibnamefont
  {Zabolotnyy}}, \bibinfo {author} {\bibfnamefont {A.}~\bibnamefont
  {Koitzsch}}, \bibinfo {author} {\bibfnamefont {M.}~\bibnamefont {Knupfer}},
  \bibinfo {author} {\bibfnamefont {B.}~\bibnamefont {B\"uchner}}, \bibinfo
  {author} {\bibfnamefont {M.~S.}\ \bibnamefont {Viazovska}}, \bibinfo {author}
  {\bibfnamefont {G.~L.}\ \bibnamefont {Sun}}, \bibinfo {author} {\bibfnamefont
  {V.}~\bibnamefont {Hinkov}}, \bibinfo {author} {\bibfnamefont {A.~V.}\
  \bibnamefont {Boris}}, \bibinfo {author} {\bibfnamefont {C.~T.}\ \bibnamefont
  {Lin}}, \bibinfo {author} {\bibfnamefont {B.}~\bibnamefont {Keimer}},
  \bibinfo {author} {\bibfnamefont {A.}~\bibnamefont {Varykhalov}}, \bibinfo
  {author} {\bibfnamefont {A.~A.}\ \bibnamefont {Kordyuk}}, \ and\ \bibinfo
  {author} {\bibfnamefont {S.~V.}\ \bibnamefont {Borisenko}},\ }\href {\doibase
  10.1103/PhysRevB.79.054517} {\bibfield  {journal} {\bibinfo  {journal} {Phys.
  Rev. B}\ }\textbf {\bibinfo {volume} {79}},\ \bibinfo {eid} {054517}
  (\bibinfo {year} {2009})}\BibitemShut {NoStop}%
\bibitem [{\citenamefont {Scalapino}\ and\ \citenamefont
  {Devereaux}(2009)}]{Scalapino:2009}%
  \BibitemOpen
  \bibfield  {author} {\bibinfo {author} {\bibfnamefont {D.~J.}\ \bibnamefont
  {Scalapino}}\ and\ \bibinfo {author} {\bibfnamefont {T.~P.}\ \bibnamefont
  {Devereaux}},\ }\href {\doibase 10.1103/PhysRevB.80.140512} {\bibfield
  {journal} {\bibinfo  {journal} {Phys. Rev. B}\ }\textbf {\bibinfo {volume}
  {80}},\ \bibinfo {eid} {140512} (\bibinfo {year} {2009})}\BibitemShut
  {NoStop}%
\bibitem [{\citenamefont {Scalapino}(2012)}]{Scalapino:2012}%
  \BibitemOpen
  \bibfield  {author} {\bibinfo {author} {\bibfnamefont {D.~J.}\ \bibnamefont
  {Scalapino}},\ }\href {\doibase 10.1103/RevModPhys.84.1383} {\bibfield
  {journal} {\bibinfo  {journal} {Rev. Mod. Phys.}\ }\textbf {\bibinfo {volume}
  {84}},\ \bibinfo {pages} {1383} (\bibinfo {year} {2012})}\BibitemShut
  {NoStop}%
\bibitem [{\citenamefont {Choi}\ \emph {et~al.}(2010)\citenamefont {Choi},
  \citenamefont {Lemmens}, \citenamefont {Eremin}, \citenamefont {Zwicknagl},
  \citenamefont {Berger}, \citenamefont {Sun}, \citenamefont {Sun},\ and\
  \citenamefont {Lin}}]{Choi:2010}%
  \BibitemOpen
  \bibfield  {author} {\bibinfo {author} {\bibfnamefont {K.-Y.}\ \bibnamefont
  {Choi}}, \bibinfo {author} {\bibfnamefont {P.}~\bibnamefont {Lemmens}},
  \bibinfo {author} {\bibfnamefont {I.}~\bibnamefont {Eremin}}, \bibinfo
  {author} {\bibfnamefont {G.}~\bibnamefont {Zwicknagl}}, \bibinfo {author}
  {\bibfnamefont {H.}~\bibnamefont {Berger}}, \bibinfo {author} {\bibfnamefont
  {G.~L.}\ \bibnamefont {Sun}}, \bibinfo {author} {\bibfnamefont {D.~L.}\
  \bibnamefont {Sun}}, \ and\ \bibinfo {author} {\bibfnamefont {C.~T.}\
  \bibnamefont {Lin}},\ }\href {\doibase 10.1088/0953-8984/22/11/115802}
  {\bibfield  {journal} {\bibinfo  {journal} {J. Phys.: Condens. Matter}\
  }\textbf {\bibinfo {volume} {22}},\ \bibinfo {pages} {115802} (\bibinfo
  {year} {2010})}\BibitemShut {NoStop}%
\bibitem [{\citenamefont {Gallais}\ \emph {et~al.}(2013)\citenamefont
  {Gallais}, \citenamefont {Fernandes}, \citenamefont {Paul}, \citenamefont
  {Chauvi\`ere}, \citenamefont {Yang}, \citenamefont {M\'easson}, \citenamefont
  {Cazayous}, \citenamefont {Sacuto}, \citenamefont {Colson},\ and\
  \citenamefont {Forget}}]{Gallais:2013}%
  \BibitemOpen
  \bibfield  {author} {\bibinfo {author} {\bibfnamefont {Y.}~\bibnamefont
  {Gallais}}, \bibinfo {author} {\bibfnamefont {R.~M.}\ \bibnamefont
  {Fernandes}}, \bibinfo {author} {\bibfnamefont {I.}~\bibnamefont {Paul}},
  \bibinfo {author} {\bibfnamefont {L.}~\bibnamefont {Chauvi\`ere}}, \bibinfo
  {author} {\bibfnamefont {Y.-X.}\ \bibnamefont {Yang}}, \bibinfo {author}
  {\bibfnamefont {M.-A.}\ \bibnamefont {M\'easson}}, \bibinfo {author}
  {\bibfnamefont {M.}~\bibnamefont {Cazayous}}, \bibinfo {author}
  {\bibfnamefont {A.}~\bibnamefont {Sacuto}}, \bibinfo {author} {\bibfnamefont
  {D.}~\bibnamefont {Colson}}, \ and\ \bibinfo {author} {\bibfnamefont
  {A.}~\bibnamefont {Forget}},\ }\href {\doibase
  10.1103/PhysRevLett.111.267001} {\bibfield  {journal} {\bibinfo  {journal}
  {Phys. Rev. Lett.}\ }\textbf {\bibinfo {volume} {111}},\ \bibinfo {pages}
  {267001} (\bibinfo {year} {2013})}\BibitemShut {NoStop}%
\bibitem [{\citenamefont {B{\"o}hmer}\ \emph {et~al.}(2014)\citenamefont
  {B{\"o}hmer}, \citenamefont {Burger}, \citenamefont {Hardy}, \citenamefont
  {Wolf}, \citenamefont {Schweiss}, \citenamefont {Fromknecht}, \citenamefont
  {Reinecker}, \citenamefont {Schranz},\ and\ \citenamefont
  {Meingast}}]{Bohmer:2014}%
  \BibitemOpen
  \bibfield  {author} {\bibinfo {author} {\bibfnamefont {A.~E.}\ \bibnamefont
  {B{\"o}hmer}}, \bibinfo {author} {\bibfnamefont {P.}~\bibnamefont {Burger}},
  \bibinfo {author} {\bibfnamefont {F.}~\bibnamefont {Hardy}}, \bibinfo
  {author} {\bibfnamefont {T.}~\bibnamefont {Wolf}}, \bibinfo {author}
  {\bibfnamefont {P.}~\bibnamefont {Schweiss}}, \bibinfo {author}
  {\bibfnamefont {R.}~\bibnamefont {Fromknecht}}, \bibinfo {author}
  {\bibfnamefont {M.}~\bibnamefont {Reinecker}}, \bibinfo {author}
  {\bibfnamefont {W.}~\bibnamefont {Schranz}}, \ and\ \bibinfo {author}
  {\bibfnamefont {C.}~\bibnamefont {Meingast}},\ }\href {\doibase
  10.1103/PhysRevLett.112.047001} {\bibfield  {journal} {\bibinfo  {journal}
  {Phys. Rev. Lett.}\ }\textbf {\bibinfo {volume} {112}},\ \bibinfo {pages}
  {047001} (\bibinfo {year} {2014})}\BibitemShut {NoStop}%
\bibitem [{\citenamefont {Shen}\ \emph {et~al.}(2011)\citenamefont {Shen},
  \citenamefont {Yang}, \citenamefont {Wang}, \citenamefont {Han},
  \citenamefont {Zeng}, \citenamefont {Shan}, \citenamefont {Cong},\ and\
  \citenamefont {Wen}}]{Shen:2011}%
  \BibitemOpen
  \bibfield  {author} {\bibinfo {author} {\bibfnamefont {B.}~\bibnamefont
  {Shen}}, \bibinfo {author} {\bibfnamefont {H.}~\bibnamefont {Yang}}, \bibinfo
  {author} {\bibfnamefont {Z.-S.}\ \bibnamefont {Wang}}, \bibinfo {author}
  {\bibfnamefont {F.}~\bibnamefont {Han}}, \bibinfo {author} {\bibfnamefont
  {B.}~\bibnamefont {Zeng}}, \bibinfo {author} {\bibfnamefont {L.}~\bibnamefont
  {Shan}}, \bibinfo {author} {\bibfnamefont {R.}~\bibnamefont {Cong}}, \ and\
  \bibinfo {author} {\bibfnamefont {H.-H.}\ \bibnamefont {Wen}},\ }\href
  {\doibase 10.1103/PhysRevB.84.184512} {\bibfield  {journal} {\bibinfo
  {journal} {Phys. Rev. B}\ }\textbf {\bibinfo {volume} {84}},\ \bibinfo
  {pages} {184512} (\bibinfo {year} {2011})}\BibitemShut {NoStop}%
\bibitem [{\citenamefont {Karkin}\ \emph {et~al.}(2014)\citenamefont {Karkin},
  \citenamefont {Wolf},\ and\ \citenamefont {Goshchitskii}}]{Karkin:2014}%
  \BibitemOpen
  \bibfield  {author} {\bibinfo {author} {\bibfnamefont {A.~E.}\ \bibnamefont
  {Karkin}}, \bibinfo {author} {\bibfnamefont {T.}~\bibnamefont {Wolf}}, \ and\
  \bibinfo {author} {\bibfnamefont {B.~N.}\ \bibnamefont {Goshchitskii}},\
  }\href {\doibase 10.1088/0953-8984/26/27/275702} {\bibfield  {journal}
  {\bibinfo  {journal} {J. Phys.: Condens. Matter}\ }\textbf {\bibinfo {volume}
  {26}},\ \bibinfo {pages} {275702} (\bibinfo {year} {2014})}\BibitemShut
  {NoStop}%
\bibitem [{\citenamefont {Chu}\ \emph {et~al.}(2009)\citenamefont {Chu},
  \citenamefont {Analytis}, \citenamefont {Kucharczyk},\ and\ \citenamefont
  {Fisher}}]{Chu:2009}%
  \BibitemOpen
  \bibfield  {author} {\bibinfo {author} {\bibfnamefont {J.-H.}\ \bibnamefont
  {Chu}}, \bibinfo {author} {\bibfnamefont {J.~G.}\ \bibnamefont {Analytis}},
  \bibinfo {author} {\bibfnamefont {C.}~\bibnamefont {Kucharczyk}}, \ and\
  \bibinfo {author} {\bibfnamefont {I.~R.}\ \bibnamefont {Fisher}},\ }\href
  {\doibase 10.1103/PhysRevB.79.014506} {\bibfield  {journal} {\bibinfo
  {journal} {Phys. Rev. B}\ }\textbf {\bibinfo {volume} {79}},\ \bibinfo {eid}
  {014506} (\bibinfo {year} {2009})}\BibitemShut {NoStop}%
\bibitem [{\citenamefont {Shastry}\ and\ \citenamefont
  {Shraiman}(1990)}]{Shastry:1990}%
  \BibitemOpen
  \bibfield  {author} {\bibinfo {author} {\bibfnamefont {B.~S.}\ \bibnamefont
  {Shastry}}\ and\ \bibinfo {author} {\bibfnamefont {B.~I.}\ \bibnamefont
  {Shraiman}},\ }\href {\doibase 10.1103/PhysRevLett.65.1068} {\bibfield
  {journal} {\bibinfo  {journal} {Phys. Rev. Lett.}\ }\textbf {\bibinfo
  {volume} {65}},\ \bibinfo {pages} {1068} (\bibinfo {year}
  {1990})}\BibitemShut {NoStop}%
\bibitem [{\citenamefont {Opel}\ \emph {et~al.}(2000)\citenamefont {Opel},
  \citenamefont {Nemetschek}, \citenamefont {Hoffmann}, \citenamefont
  {Philipp}, \citenamefont {M\"uller}, \citenamefont {Hackl}, \citenamefont
  {T\"utt\ifmmode~\mbox{\H{o}}\else \H{o}\fi{}}, \citenamefont {Erb},
  \citenamefont {Revaz}, \citenamefont {Walker}, \citenamefont {Berger},\ and\
  \citenamefont {Forr\'o}}]{Opel:2000}%
  \BibitemOpen
  \bibfield  {author} {\bibinfo {author} {\bibfnamefont {M.}~\bibnamefont
  {Opel}}, \bibinfo {author} {\bibfnamefont {R.}~\bibnamefont {Nemetschek}},
  \bibinfo {author} {\bibfnamefont {C.}~\bibnamefont {Hoffmann}}, \bibinfo
  {author} {\bibfnamefont {R.}~\bibnamefont {Philipp}}, \bibinfo {author}
  {\bibfnamefont {P.~F.}\ \bibnamefont {M\"uller}}, \bibinfo {author}
  {\bibfnamefont {R.}~\bibnamefont {Hackl}}, \bibinfo {author} {\bibfnamefont
  {I.}~\bibnamefont {T\"utt\ifmmode~\mbox{\H{o}}\else \H{o}\fi{}}}, \bibinfo
  {author} {\bibfnamefont {A.}~\bibnamefont {Erb}}, \bibinfo {author}
  {\bibfnamefont {B.}~\bibnamefont {Revaz}}, \bibinfo {author} {\bibfnamefont
  {E.}~\bibnamefont {Walker}}, \bibinfo {author} {\bibfnamefont
  {H.}~\bibnamefont {Berger}}, \ and\ \bibinfo {author} {\bibfnamefont
  {L.}~\bibnamefont {Forr\'o}},\ }\href {\doibase 10.1103/PhysRevB.61.9752}
  {\bibfield  {journal} {\bibinfo  {journal} {Phys. Rev. B}\ }\textbf {\bibinfo
  {volume} {61}},\ \bibinfo {pages} {9752} (\bibinfo {year}
  {2000})}\BibitemShut {NoStop}%
\bibitem [{\citenamefont {Devereaux}\ and\ \citenamefont
  {Einzel}(1995)}]{Devereaux:1995a}%
  \BibitemOpen
  \bibfield  {author} {\bibinfo {author} {\bibfnamefont {T.~P.}\ \bibnamefont
  {Devereaux}}\ and\ \bibinfo {author} {\bibfnamefont {D.}~\bibnamefont
  {Einzel}},\ }\href {\doibase 10.1103/PhysRevB.51.16336} {\bibfield  {journal}
  {\bibinfo  {journal} {Phys. Rev. B}\ }\textbf {\bibinfo {volume} {51}},\
  \bibinfo {pages} {16336} (\bibinfo {year} {1995})}\BibitemShut {NoStop}%
\bibitem [{\citenamefont {Kosztin}\ and\ \citenamefont
  {Zawadowski}(1991)}]{Kosztin:1991}%
  \BibitemOpen
  \bibfield  {author} {\bibinfo {author} {\bibfnamefont {J.}~\bibnamefont
  {Kosztin}}\ and\ \bibinfo {author} {\bibfnamefont {A.}~\bibnamefont
  {Zawadowski}},\ }\href {\doibase 10.1016/0038-1098(91)90123-D} {\bibfield
  {journal} {\bibinfo  {journal} {Solid State Commun.}\ }\textbf {\bibinfo
  {volume} {78}},\ \bibinfo {pages} {1029} (\bibinfo {year}
  {1991})}\BibitemShut {NoStop}%
\bibitem [{\citenamefont {Freericks}\ \emph {et~al.}(2005)\citenamefont
  {Freericks}, \citenamefont {Devereaux}, \citenamefont {Moraghebi},\ and\
  \citenamefont {L.}}]{Freericks:2005}%
  \BibitemOpen
  \bibfield  {author} {\bibinfo {author} {\bibfnamefont {J.~K.}\ \bibnamefont
  {Freericks}}, \bibinfo {author} {\bibfnamefont {T.~P.}\ \bibnamefont
  {Devereaux}}, \bibinfo {author} {\bibfnamefont {M.}~\bibnamefont
  {Moraghebi}}, \ and\ \bibinfo {author} {\bibfnamefont {C.~S.}\ \bibnamefont
  {L.}},\ }\href {\doibase 10.1103/PhysRevLett.94.216401} {\bibfield  {journal}
  {\bibinfo  {journal} {Phys. Rev. Lett.}\ }\textbf {\bibinfo {volume} {94}},\
  \bibinfo {pages} {216401} (\bibinfo {year} {2005})}\BibitemShut {NoStop}%
\bibitem [{\citenamefont {Zawadowski}\ and\ \citenamefont
  {Cardona}(1990)}]{Zawadowski:1990}%
  \BibitemOpen
  \bibfield  {author} {\bibinfo {author} {\bibfnamefont {A.}~\bibnamefont
  {Zawadowski}}\ and\ \bibinfo {author} {\bibfnamefont {M.}~\bibnamefont
  {Cardona}},\ }\href {\doibase 10.1103/PhysRevB.42.10732} {\bibfield
  {journal} {\bibinfo  {journal} {Phys. Rev. B}\ }\textbf {\bibinfo {volume}
  {42}},\ \bibinfo {pages} {10732} (\bibinfo {year} {1990})}\BibitemShut
  {NoStop}%
\bibitem [{\citenamefont {Devereaux}(2003)}]{Devereaux:2003}%
  \BibitemOpen
  \bibfield  {author} {\bibinfo {author} {\bibfnamefont {T.~P.}\ \bibnamefont
  {Devereaux}},\ }\href {\doibase 10.1103/PhysRevB.68.094503} {\bibfield
  {journal} {\bibinfo  {journal} {Phys. Rev. B}\ }\textbf {\bibinfo {volume}
  {68}},\ \bibinfo {pages} {094503} (\bibinfo {year} {2003})}\BibitemShut
  {NoStop}%
\bibitem [{\citenamefont {Devereaux}\ and\ \citenamefont
  {Hackl}(2007)}]{Devereaux:2007}%
  \BibitemOpen
  \bibfield  {author} {\bibinfo {author} {\bibfnamefont {T.~P.}\ \bibnamefont
  {Devereaux}}\ and\ \bibinfo {author} {\bibfnamefont {R.}~\bibnamefont
  {Hackl}},\ }\href {\doibase 10.1103/RevModPhys.79.175} {\bibfield  {journal}
  {\bibinfo  {journal} {Rev. Mod. Phys.}\ }\textbf {\bibinfo {volume} {79}},\
  \bibinfo {pages} {175} (\bibinfo {year} {2007})}\BibitemShut {NoStop}%
\bibitem [{\citenamefont {Opel}\ \emph {et~al.}(1999)\citenamefont {Opel},
  \citenamefont {Hackl}, \citenamefont {Devereaux}, \citenamefont {Virosztek},
  \citenamefont {Zawadowski}, \citenamefont {Erb}, \citenamefont {Walker},
  \citenamefont {Berger},\ and\ \citenamefont {Forr\'o}}]{Opel:1999b}%
  \BibitemOpen
  \bibfield  {author} {\bibinfo {author} {\bibfnamefont {M.}~\bibnamefont
  {Opel}}, \bibinfo {author} {\bibfnamefont {R.}~\bibnamefont {Hackl}},
  \bibinfo {author} {\bibfnamefont {T.~P.}\ \bibnamefont {Devereaux}}, \bibinfo
  {author} {\bibfnamefont {A.}~\bibnamefont {Virosztek}}, \bibinfo {author}
  {\bibfnamefont {A.}~\bibnamefont {Zawadowski}}, \bibinfo {author}
  {\bibfnamefont {A.}~\bibnamefont {Erb}}, \bibinfo {author} {\bibfnamefont
  {E.}~\bibnamefont {Walker}}, \bibinfo {author} {\bibfnamefont
  {H.}~\bibnamefont {Berger}}, \ and\ \bibinfo {author} {\bibfnamefont
  {L.}~\bibnamefont {Forr\'o}},\ }\href {\doibase 10.1103/PhysRevB.60.9836}
  {\bibfield  {journal} {\bibinfo  {journal} {Phys. Rev. B}\ }\textbf {\bibinfo
  {volume} {60}},\ \bibinfo {pages} {9836} (\bibinfo {year}
  {1999})}\BibitemShut {NoStop}%
\bibitem [{\citenamefont {Manske}(2004)}]{Manske:2004}%
  \BibitemOpen
  \bibfield  {author} {\bibinfo {author} {\bibfnamefont {D.}~\bibnamefont
  {Manske}},\ }\href@noop {} {\emph {\bibinfo {title} {{Theory of
  Unconventional Superconductors}}}},\ Vol.\ \bibinfo {volume} {202}\ (\bibinfo
   {publisher} {Springer Tracts in Modern Physics},\ \bibinfo {year}
  {2004})\BibitemShut {NoStop}%
\bibitem [{\citenamefont {Yoon}\ \emph {et~al.}(2000)\citenamefont {Yoon},
  \citenamefont {R\"ubhausen}, \citenamefont {Cooper}, \citenamefont {Kim},\
  and\ \citenamefont {Cheong}}]{Yoon:2000}%
  \BibitemOpen
  \bibfield  {author} {\bibinfo {author} {\bibfnamefont {S.}~\bibnamefont
  {Yoon}}, \bibinfo {author} {\bibfnamefont {M.}~\bibnamefont {R\"ubhausen}},
  \bibinfo {author} {\bibfnamefont {S.~L.}\ \bibnamefont {Cooper}}, \bibinfo
  {author} {\bibfnamefont {K.~H.}\ \bibnamefont {Kim}}, \ and\ \bibinfo
  {author} {\bibfnamefont {S.-W.}\ \bibnamefont {Cheong}},\ }\href {\doibase
  10.1103/PhysRevLett.85.3297} {\bibfield  {journal} {\bibinfo  {journal}
  {Phys. Rev. Lett.}\ }\textbf {\bibinfo {volume} {85}},\ \bibinfo {pages}
  {3297} (\bibinfo {year} {2000})}\BibitemShut {NoStop}%
\bibitem [{\citenamefont {Venturini}\ \emph {et~al.}(2002)\citenamefont
  {Venturini}, \citenamefont {Zhang}, \citenamefont {Hackl}, \citenamefont
  {Lucarelli}, \citenamefont {Lupi}, \citenamefont {Ortolani}, \citenamefont
  {Calvani}, \citenamefont {Kikugawa},\ and\ \citenamefont
  {Fujita}}]{Venturini:2002c}%
  \BibitemOpen
  \bibfield  {author} {\bibinfo {author} {\bibfnamefont {F.}~\bibnamefont
  {Venturini}}, \bibinfo {author} {\bibfnamefont {Q.-M.}\ \bibnamefont
  {Zhang}}, \bibinfo {author} {\bibfnamefont {R.}~\bibnamefont {Hackl}},
  \bibinfo {author} {\bibfnamefont {A.}~\bibnamefont {Lucarelli}}, \bibinfo
  {author} {\bibfnamefont {S.}~\bibnamefont {Lupi}}, \bibinfo {author}
  {\bibfnamefont {M.}~\bibnamefont {Ortolani}}, \bibinfo {author}
  {\bibfnamefont {P.}~\bibnamefont {Calvani}}, \bibinfo {author} {\bibfnamefont
  {N.}~\bibnamefont {Kikugawa}}, \ and\ \bibinfo {author} {\bibfnamefont
  {T.}~\bibnamefont {Fujita}},\ }\href@noop {} {\bibfield  {journal} {\bibinfo
  {journal} {Phys. Rev. B}\ }\textbf {\bibinfo {volume} {66}},\ \bibinfo
  {pages} {060502} (\bibinfo {year} {2002})}\BibitemShut {NoStop}%
\bibitem [{\citenamefont {Gallais}\ and\ \citenamefont
  {Paul}(2016)}]{Gallais:2016}%
  \BibitemOpen
  \bibfield  {author} {\bibinfo {author} {\bibfnamefont {Y.}~\bibnamefont
  {Gallais}}\ and\ \bibinfo {author} {\bibfnamefont {I.}~\bibnamefont {Paul}},\
  }\href {\doibase http://dx.doi.org/10.1016/j.crhy.2015.10.001} {\bibfield
  {journal} {\bibinfo  {journal} {C. R. Physique}\ }\textbf {\bibinfo {volume}
  {17}},\ \bibinfo {pages} {113 } (\bibinfo {year} {2016})}\BibitemShut
  {NoStop}%
\bibitem [{\citenamefont {Khodas}\ and\ \citenamefont
  {Levchenko}(2015)}]{Khodas:2015}%
  \BibitemOpen
  \bibfield  {author} {\bibinfo {author} {\bibfnamefont {M.}~\bibnamefont
  {Khodas}}\ and\ \bibinfo {author} {\bibfnamefont {A.}~\bibnamefont
  {Levchenko}},\ }\href {\doibase 10.1103/PhysRevB.91.235119} {\bibfield
  {journal} {\bibinfo  {journal} {Phys. Rev. B}\ }\textbf {\bibinfo {volume}
  {91}},\ \bibinfo {pages} {235119} (\bibinfo {year} {2015})}\BibitemShut
  {NoStop}%
\bibitem [{\citenamefont {Karahasanovic}\ \emph {et~al.}(2015)\citenamefont
  {Karahasanovic}, \citenamefont {Kretzschmar}, \citenamefont {B\"ohm},
  \citenamefont {Hackl}, \citenamefont {Paul}, \citenamefont {Gallais},\ and\
  \citenamefont {Schmalian}}]{Karahasanovic:2015}%
  \BibitemOpen
  \bibfield  {author} {\bibinfo {author} {\bibfnamefont {U.}~\bibnamefont
  {Karahasanovic}}, \bibinfo {author} {\bibfnamefont {F.}~\bibnamefont
  {Kretzschmar}}, \bibinfo {author} {\bibfnamefont {T.}~\bibnamefont {B\"ohm}},
  \bibinfo {author} {\bibfnamefont {R.}~\bibnamefont {Hackl}}, \bibinfo
  {author} {\bibfnamefont {I.}~\bibnamefont {Paul}}, \bibinfo {author}
  {\bibfnamefont {Y.}~\bibnamefont {Gallais}}, \ and\ \bibinfo {author}
  {\bibfnamefont {J.}~\bibnamefont {Schmalian}},\ }\href {\doibase
  10.1103/PhysRevB.92.075134} {\bibfield  {journal} {\bibinfo  {journal} {Phys.
  Rev. B}\ }\textbf {\bibinfo {volume} {92}},\ \bibinfo {pages} {075134}
  (\bibinfo {year} {2015})}\BibitemShut {NoStop}%
\bibitem [{\citenamefont {Yamase}\ and\ \citenamefont
  {Zeyher}(2013)}]{Yamase:2013}%
  \BibitemOpen
  \bibfield  {author} {\bibinfo {author} {\bibfnamefont {H.}~\bibnamefont
  {Yamase}}\ and\ \bibinfo {author} {\bibfnamefont {R.}~\bibnamefont
  {Zeyher}},\ }\href {\doibase 10.1103/PhysRevB.88.125120} {\bibfield
  {journal} {\bibinfo  {journal} {Phys. Rev. B}\ }\textbf {\bibinfo {volume}
  {88}},\ \bibinfo {pages} {125120} (\bibinfo {year} {2013})}\BibitemShut
  {NoStop}%
\bibitem [{\citenamefont {Yamase}\ and\ \citenamefont
  {Zeyher}(2015)}]{Yamase:2015}%
  \BibitemOpen
  \bibfield  {author} {\bibinfo {author} {\bibfnamefont {H.}~\bibnamefont
  {Yamase}}\ and\ \bibinfo {author} {\bibfnamefont {R.}~\bibnamefont
  {Zeyher}},\ }\href {\doibase 10.1088/1367-2630/17/7/073030} {\bibfield
  {journal} {\bibinfo  {journal} {New J. Phys.}\ }\textbf {\bibinfo {volume}
  {17}},\ \bibinfo {pages} {073030} (\bibinfo {year} {2015})}\BibitemShut
  {NoStop}%
\bibitem [{\citenamefont {Fernandes}\ and\ \citenamefont
  {Schmalian}(2012)}]{Fernandes:2012}%
  \BibitemOpen
  \bibfield  {author} {\bibinfo {author} {\bibfnamefont {R.~M.}\ \bibnamefont
  {Fernandes}}\ and\ \bibinfo {author} {\bibfnamefont {J.}~\bibnamefont
  {Schmalian}},\ }\href {\doibase 10.1088/0953-2048/25/8/084005} {\bibfield
  {journal} {\bibinfo  {journal} {Supercond. Sci. Technol.}\ }\textbf {\bibinfo
  {volume} {25}},\ \bibinfo {pages} {084005} (\bibinfo {year}
  {2012})}\BibitemShut {NoStop}%
\bibitem [{\citenamefont {Kontani}\ and\ \citenamefont
  {Onari}(2010)}]{Kontani:2010}%
  \BibitemOpen
  \bibfield  {author} {\bibinfo {author} {\bibfnamefont {H.}~\bibnamefont
  {Kontani}}\ and\ \bibinfo {author} {\bibfnamefont {S.}~\bibnamefont
  {Onari}},\ }\href {\doibase 10.1103/PhysRevLett.104.157001} {\bibfield
  {journal} {\bibinfo  {journal} {Phys. Rev. Lett.}\ }\textbf {\bibinfo
  {volume} {104}},\ \bibinfo {pages} {157001} (\bibinfo {year}
  {2010})}\BibitemShut {NoStop}%
\bibitem [{\citenamefont {Fernandes}\ \emph {et~al.}(2014)\citenamefont
  {Fernandes}, \citenamefont {Chubukov},\ and\ \citenamefont
  {Schmalian}}]{Fernandes:2014}%
  \BibitemOpen
  \bibfield  {author} {\bibinfo {author} {\bibfnamefont {R.~M.}\ \bibnamefont
  {Fernandes}}, \bibinfo {author} {\bibfnamefont {A.~V.}\ \bibnamefont
  {Chubukov}}, \ and\ \bibinfo {author} {\bibfnamefont {J.}~\bibnamefont
  {Schmalian}},\ }\href {\doibase 10.1038/nphys2877} {\bibfield  {journal}
  {\bibinfo  {journal} {Nature Phys.}\ }\textbf {\bibinfo {volume} {10}},\
  \bibinfo {pages} {97} (\bibinfo {year} {2014})}\BibitemShut {NoStop}%
\bibitem [{\citenamefont {Yoshizawa}\ \emph {et~al.}(2012)\citenamefont
  {Yoshizawa}, \citenamefont {Kimura}, \citenamefont {Chiba}, \citenamefont
  {Simayi}, \citenamefont {Nakanishi}, \citenamefont {Kihou}, \citenamefont
  {Lee}, \citenamefont {Iyo}, \citenamefont {Eisaki}, \citenamefont
  {Nakajima},\ and\ \citenamefont {Uchida}}]{Yoshizawa:2012}%
  \BibitemOpen
  \bibfield  {author} {\bibinfo {author} {\bibfnamefont {M.}~\bibnamefont
  {Yoshizawa}}, \bibinfo {author} {\bibfnamefont {D.}~\bibnamefont {Kimura}},
  \bibinfo {author} {\bibfnamefont {T.}~\bibnamefont {Chiba}}, \bibinfo
  {author} {\bibfnamefont {S.}~\bibnamefont {Simayi}}, \bibinfo {author}
  {\bibfnamefont {Y.}~\bibnamefont {Nakanishi}}, \bibinfo {author}
  {\bibfnamefont {K.}~\bibnamefont {Kihou}}, \bibinfo {author} {\bibfnamefont
  {C.-H.}\ \bibnamefont {Lee}}, \bibinfo {author} {\bibfnamefont
  {A.}~\bibnamefont {Iyo}}, \bibinfo {author} {\bibfnamefont {H.}~\bibnamefont
  {Eisaki}}, \bibinfo {author} {\bibfnamefont {M.}~\bibnamefont {Nakajima}}, \
  and\ \bibinfo {author} {\bibfnamefont {S.-i.}\ \bibnamefont {Uchida}},\
  }\href {\doibase 10.1143/JPSJ.81.024604} {\bibfield  {journal} {\bibinfo
  {journal} {J. Phys. Soc. Japan}\ }\textbf {\bibinfo {volume} {81}},\ \bibinfo
  {pages} {024604} (\bibinfo {year} {2012})}\BibitemShut {NoStop}%
\bibitem [{\citenamefont {B{\"o}hmer}\ and\ \citenamefont
  {Meingast}(2016)}]{Bohmer:2016}%
  \BibitemOpen
  \bibfield  {author} {\bibinfo {author} {\bibfnamefont {A.~E.}\ \bibnamefont
  {B{\"o}hmer}}\ and\ \bibinfo {author} {\bibfnamefont {C.}~\bibnamefont
  {Meingast}},\ }\href {\doibase http://dx.doi.org/10.1016/j.crhy.2015.07.001}
  {\bibfield  {journal} {\bibinfo  {journal} {C. R. Physique}\ }\textbf
  {\bibinfo {volume} {17}},\ \bibinfo {pages} {90 } (\bibinfo {year}
  {2016})}\BibitemShut {NoStop}%
\bibitem [{\citenamefont {Cano}\ \emph {et~al.}(2010)\citenamefont {Cano},
  \citenamefont {Civelli}, \citenamefont {Eremin},\ and\ \citenamefont
  {Paul}}]{Cano:2010}%
  \BibitemOpen
  \bibfield  {author} {\bibinfo {author} {\bibfnamefont {A.}~\bibnamefont
  {Cano}}, \bibinfo {author} {\bibfnamefont {M.}~\bibnamefont {Civelli}},
  \bibinfo {author} {\bibfnamefont {I.}~\bibnamefont {Eremin}}, \ and\ \bibinfo
  {author} {\bibfnamefont {I.}~\bibnamefont {Paul}},\ }\href {\doibase
  10.1103/PhysRevB.82.020408} {\bibfield  {journal} {\bibinfo  {journal} {Phys.
  Rev. B}\ }\textbf {\bibinfo {volume} {82}},\ \bibinfo {pages} {020408}
  (\bibinfo {year} {2010})}\BibitemShut {NoStop}%
\bibitem [{\citenamefont {Fernandes}\ and\ \citenamefont
  {Schmalian}(2010)}]{Fernandes:2010}%
  \BibitemOpen
  \bibfield  {author} {\bibinfo {author} {\bibfnamefont {R.~M.}\ \bibnamefont
  {Fernandes}}\ and\ \bibinfo {author} {\bibfnamefont {J.}~\bibnamefont
  {Schmalian}},\ }\href {\doibase 10.1103/PhysRevB.82.014521} {\bibfield
  {journal} {\bibinfo  {journal} {Phys. Rev. B}\ }\textbf {\bibinfo {volume}
  {82}},\ \bibinfo {pages} {014521} (\bibinfo {year} {2010})}\BibitemShut
  {NoStop}%
\bibitem [{\citenamefont {Aslamasov}\ and\ \citenamefont
  {Larkin}(1968)}]{Aslamasov:1968}%
  \BibitemOpen
  \bibfield  {author} {\bibinfo {author} {\bibfnamefont {L.}~\bibnamefont
  {Aslamasov}}\ and\ \bibinfo {author} {\bibfnamefont {A.}~\bibnamefont
  {Larkin}},\ }\href
  {http://www.sciencedirect.com/science/article/B6TVM-46X9PHW-GD/1/658b80956b0c460776f2d08ac9b04208}
  {\bibfield  {journal} {\bibinfo  {journal} {Phys. Rev. A}\ }\textbf {\bibinfo
  {volume} {26}},\ \bibinfo {pages} {238} (\bibinfo {year} {1968})}\BibitemShut
  {NoStop}%
\bibitem [{\citenamefont {Devereaux}(1992)}]{Devereaux:1992}%
  \BibitemOpen
  \bibfield  {author} {\bibinfo {author} {\bibfnamefont {T.~P.}\ \bibnamefont
  {Devereaux}},\ }\href {\doibase 10.1103/PhysRevB.45.12965} {\bibfield
  {journal} {\bibinfo  {journal} {Phys. Rev. B}\ }\textbf {\bibinfo {volume}
  {45}},\ \bibinfo {pages} {12965} (\bibinfo {year} {1992})}\BibitemShut
  {NoStop}%
\bibitem [{\citenamefont {Devereaux}(1995)}]{Devereaux:1995}%
  \BibitemOpen
  \bibfield  {author} {\bibinfo {author} {\bibfnamefont {T.~P.}\ \bibnamefont
  {Devereaux}},\ }\href {\doibase 10.1103/PhysRevLett.74.4313} {\bibfield
  {journal} {\bibinfo  {journal} {Phys. Rev. Lett.}\ }\textbf {\bibinfo
  {volume} {74}},\ \bibinfo {pages} {4313} (\bibinfo {year}
  {1995})}\BibitemShut {NoStop}%
\bibitem [{\citenamefont {Inosov}(2016)}]{Inosov:2016}%
  \BibitemOpen
  \bibfield  {author} {\bibinfo {author} {\bibfnamefont {D.~S.}\ \bibnamefont
  {Inosov}},\ }\href {\doibase http://dx.doi.org/10.1016/j.crhy.2015.03.001}
  {\bibfield  {journal} {\bibinfo  {journal} {C. R. Physique}\ }\textbf
  {\bibinfo {volume} {17}},\ \bibinfo {pages} {60 } (\bibinfo {year}
  {2016})}\BibitemShut {NoStop}%
\bibitem [{\citenamefont {Devereaux}(1993)}]{Devereaux:1993}%
  \BibitemOpen
  \bibfield  {author} {\bibinfo {author} {\bibfnamefont {T.~P.}\ \bibnamefont
  {Devereaux}},\ }\href {\doibase 10.1103/PhysRevB.47.5230} {\bibfield
  {journal} {\bibinfo  {journal} {Phys. Rev. B}\ }\textbf {\bibinfo {volume}
  {47}},\ \bibinfo {pages} {5230} (\bibinfo {year} {1993})}\BibitemShut
  {NoStop}%
\bibitem [{\citenamefont {Christianson}\ \emph {et~al.}(2008)\citenamefont
  {Christianson}, \citenamefont {Goremychkin}, \citenamefont {Osborn},
  \citenamefont {Rosenkranz}, \citenamefont {Lumsden}, \citenamefont
  {Malliakas}, \citenamefont {Todorov}, \citenamefont {Claus}, \citenamefont
  {Chung}, \citenamefont {Kanatzidis}, \citenamefont {Bewley},\ and\
  \citenamefont {Guidi}}]{Christianson:2008}%
  \BibitemOpen
  \bibfield  {author} {\bibinfo {author} {\bibfnamefont {A.~D.}\ \bibnamefont
  {Christianson}}, \bibinfo {author} {\bibfnamefont {E.~A.}\ \bibnamefont
  {Goremychkin}}, \bibinfo {author} {\bibfnamefont {R.}~\bibnamefont {Osborn}},
  \bibinfo {author} {\bibfnamefont {S.}~\bibnamefont {Rosenkranz}}, \bibinfo
  {author} {\bibfnamefont {M.~D.}\ \bibnamefont {Lumsden}}, \bibinfo {author}
  {\bibfnamefont {C.~D.}\ \bibnamefont {Malliakas}}, \bibinfo {author}
  {\bibfnamefont {I.~S.}\ \bibnamefont {Todorov}}, \bibinfo {author}
  {\bibfnamefont {H.}~\bibnamefont {Claus}}, \bibinfo {author} {\bibfnamefont
  {D.~Y.}\ \bibnamefont {Chung}}, \bibinfo {author} {\bibfnamefont {M.~G.}\
  \bibnamefont {Kanatzidis}}, \bibinfo {author} {\bibfnamefont {R.~I.}\
  \bibnamefont {Bewley}}, \ and\ \bibinfo {author} {\bibfnamefont
  {T.}~\bibnamefont {Guidi}},\ }\href {\doibase doi:10.1038/nature07625}
  {\bibfield  {journal} {\bibinfo  {journal} {Nature}\ }\textbf {\bibinfo
  {volume} {456}},\ \bibinfo {pages} {930} (\bibinfo {year}
  {2008})}\BibitemShut {NoStop}%
\end{thebibliography}
%
%
\end{document}